\documentclass[preprint,pre,aps,11pt]{revtex4}
\usepackage{graphicx}
\usepackage{float}
\usepackage{dcolumn}
\usepackage{amsmath,amssymb}
\usepackage{boxedminipage}
\usepackage{array}
\usepackage{flafter}
\usepackage[mathscr]{euscript}
\usepackage[tight,TABTOPCAP]{subfigure}
\usepackage[normalsize]{caption}
\usepackage{setspace}
\usepackage{verbatim}
\newcolumntype{.}{D{.}{.}{1}}
\usepackage{bm}
\usepackage[utf8]{inputenc}
\usepackage[T1]{fontenc}
\usepackage{mathptmx}
\begin{document}
\title{Light propagation in 2D and 3D slabs of reflective colloidal particles in solution: 
the effect of interfaces and inter-particle correlations}
\author{Raffaela Cabriolu,$^{(1) \dagger}$, Sarah Dungan,$^{(2)}$, and Pietro Ballone$^{(2,3)}$}
\affiliation{(1) Department of Physics, Norwegian University of Science and Technology (NTNU), 7491 Trondheim}
\affiliation{(2) School of Physics, University College, Dublin, Ireland}
\affiliation{(3) Conway Institute for Biomolecular and Biomedical Research, University College, Dublin, Ireland}
\begin{abstract}
\label{abstract}
The propagation of light across 2D and 3D slabs of reflective colloidal particles in a fluid-like state has been investigated by simulation.
The colloids are represented as hard spheres with and without an attractive square-well tail. Representative configurations of particles have 
been generated by Monte Carlo. The path of rays entering the slab normal to its planar surface has been determined by exact geometric 
scattering conditions, assuming that particles are macroscopic spheres fully reflective at the surface of their hard-core potential. The 
analysis of light paths provides the transmission and reflection coefficients, the mean-free path, the average length of transmitted and 
reflected paths, the distribution of scattering events across the slab, the angular spread of the outcoming rays as a function of 
dimensionality and thermodynamic state. The results highlight the presence of a sizeable population of very long paths, which play an 
important role in random lasing from solutions of metal particles in an optically active fluid. The output power spectrum resulting 
from the stimulated emission amplification decays asymptotically as an inverse power law. The present study goes beyond the standard approach 
based on a random walk confined between two planar interfaces and parametrised in terms of the mean-free path and scattering matrix. Here, 
instead, the mean free path, the correlation among scattering events and memory effects are not assumed a priori, but emerge from the 
underlying statistical mechanics model of interacting particles. In this way, the dependence of properties on the thermodynamic state, the 
effect of particle-particle and particle-interface correlations, of spatial inhomogeneity and memory effects are accounted for in a 
transparent way. Moreover, the approach joins smoothly the ballistic regime of light propagation at low density with the diffusive regime at 
high density of scattering centres. These properties are exploited to investigate the effect of weak polydispersivity and of large density 
fluctuations at the critical point of the model with the attractive potential tail.
\vskip 1.3truecm
{\bf $\dagger$ Corresponding author:}    Raffaela Cabriolu \ \ \ \ \ \ \ \  e-mail: raffaela.cabriolu@ntnu.no
\end{abstract}
\maketitle

\section{Introduction}
The scattering of light in disordered media is a subject of great conceptual and practical interest, which underlies natural phenomena such as 
the rainbow \cite{rainbow} and the metallic hues of selected butterflies and beetles, as well as the iridescent glazing of ancient Chinese
pottery. The interest in this subject has been renewed and enhances by its applications in photonics and nanophotonics in particular.

One specific case of application is represented by random lasing. Conventional lasers \cite{lasers} rely on the amplification of light by 
stimulated emission in an active medium 
whose population inversion is driven by a secondary light source. Their gain is typically due to the coupling to a high quality optical 
cavity, whose mode selectivity and sizeable energy density are instrumental in achieving strong beam intensity, coherence and collimation.
High amplification of light, however, does not strictly require a highly tuned optical cavity \cite{ambar, leto}, since it can be achieved 
also using simpler set-ups such as the variety of devices known as random lasers (RL) \cite{wiersma, cao}. In these systems, amplification 
relies on the propagation through an optically active medium of a beam of light whose path is lengthened by multiple reflections from a 
static or dynamic disordered distribution of scattering centres. The appeal of random lasers is due, first of all, to the simplicity of their 
fabrication, since these systems are largely self-assembled. To a lesser extent, it is also due to the contrast of structural randomness and 
cooperative optical emission. Depending on the type and structure of the underlying medium, on the feedback mechanism, and also on the light 
intensity, the output of RLs can be coherent \cite{cohe} or, more often, incoherent light \cite{incohe}.  Even random lasers can support 
discrete modes, but often, especially in the incoherent feedback case, the emission spectrum is rather broad, and directionality is lacking. 
Both features could represent advantages or disadvantages depending on the application.

Arguably, the simplest realisation of RL is represented by particles of $\mu$m to mm diameter dispersed at equilibrium conditions (colloids) 
\cite{first}, or out of equilibrium (shaken granular systems) \cite{viola, viola2} in an optically active fluid. A paradigmatic case
is represented by $\mu$m TiO$_2$ colloidal particles dissolved in a solution of fluorescent organic molecules \cite{first}. Then, the 
scattering of light is due to the difference of refraction index between particles and the solution, while the amplification is due to the 
optical pumping of the organic dye. Other dielectric oxides (ZnO, Al$_2$O$_3$, etc.) and metal particles are used as well. 

From the conceptual point of view, random lasers are at the crossroad of several different research topics. First of all, the propagation of 
wave-like excitations in a medium sprinkled by a disordered distribution of scattering centres is a central theme of statistical mechanics 
\cite{waves, car}, with implications for a broad range of subjects, exemplified by the propagation of light in glasses and defective crystals 
\cite{light} and of electrons in disordered metal alloys \cite{and, electrons1, electrons2}, the localisation of vibrations in amorphous 
materials \cite{vibr}, the scattering of light from aerosols in the atmosphere \cite{aero}, the diffusion of cold atoms in 
disordered lattices \cite{atoms}. Despite lacking some of the characteristic features of conventional lasers, also random lasers find application in a 
number of different fields, such as photonics \cite{phot}, display technology, sensing, speckle-free imaging \cite{speck}, and light therapy 
while further applications are under development in diagnostics exploiting the random lasing in human tissues infiltrated with an organic dye 
solution \cite{human}. New conceptual developments, and the introduction of new materials \cite{newmat} can still greatly extend the application
range of RLs.

Because of their broad interest, RLs have been extensively investigated by theoretical methods \cite{theo1, theo2, theo3, theo4, theo5, theo6,
theo7} and by simulation \cite{aero, simu1, levy, simu2, simu3}, accounting for a variety of aspects in the nature of the scattering centres 
and of the scattering process (Mie scattering, reflective hard spheres for colloids, ..), including the competition and sometimes the 
interference of different wave beams propagating in the medium. In the simplest case of colloids in an optically active solvent, and also with
RLs based on granular matter, theoretical and computational studies focused primarily on systems with a high density of scattering centres, in
which the propagation of light is diffusive. In this regime, the underlying kinetics of light has been, in most cases, modelled as a random 
walk in a finite slab, with preassigned 
distribution of scattering probability. More first-principle approaches to account for correlation among scattering events have been proposed
(see, for instance, Ref.~\cite{conti}) but rarely applied. The idealised random walk approaches, however, require additional parameters 
and ad hoc assumptions to model changes in the thermodynamic conditions and in the inter-particle potential, or the effect of inhomogeneities 
in the system. Thus, these parametrised models might not reproduce exactly all the correlations found in real colloidal systems, nor account 
consistently for density oscillations at surfaces and interfaces, or for large scale inhomogeneities in samples approaching criticality.

To cover also these phenomena, we model the scattering of a light beam from a collection of hard spheres with and without an attractive tail,
floating into a fluid environment, represented as a slab confined by two planar, parallel surfaces. The model closely corresponds to 
macroscopic dielectric \cite{die1, die2} or metal particles \cite{cohe, metal2} in a finite container, which could represent 
the core of a random laser. 

Particle configurations are generated by Monte Carlo simulation, hence they fully account for the exact 
correlations in the positions of scattering centres, and also they contain exactly the effect of interfaces and of density fluctuations at 
all wavelengths.

\begin{figure}[!htb]
\begin{minipage}[c]{\textwidth}
\vskip 0.7truecm
\begin{center}
\includegraphics[scale=0.60]{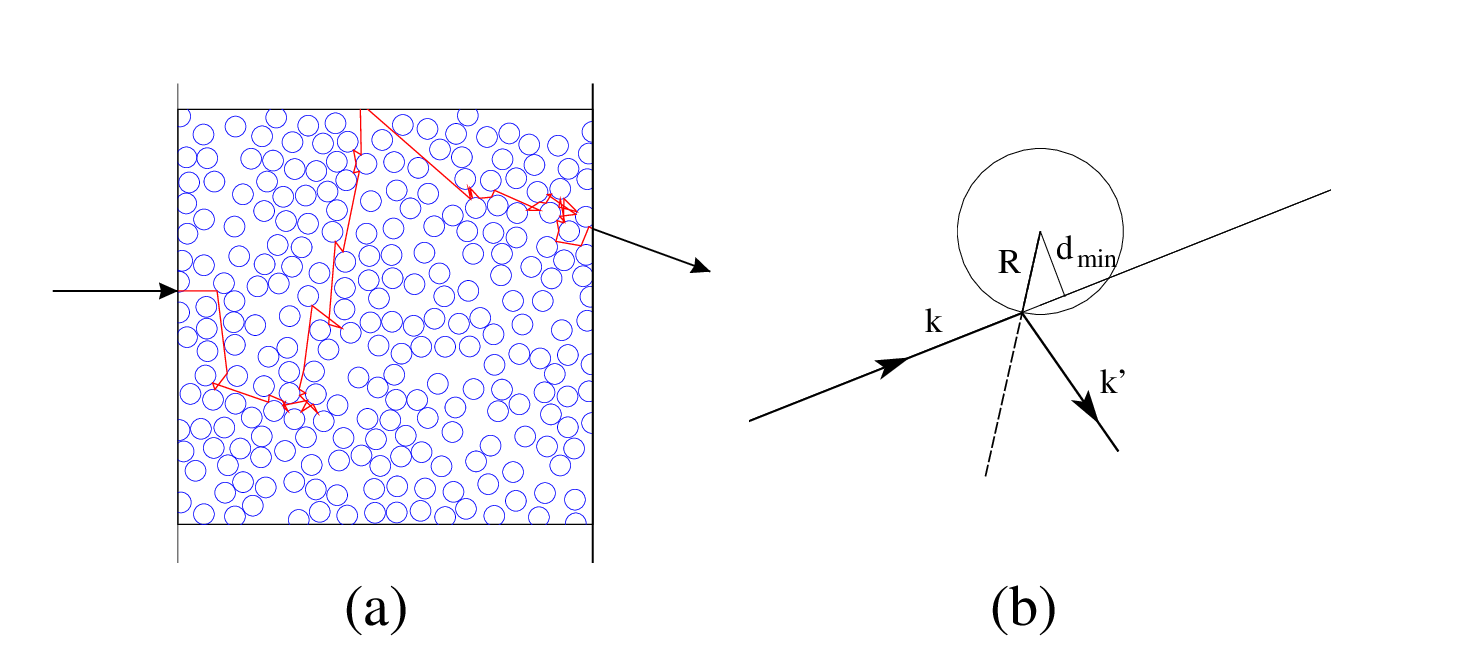}
\caption{Schematic representation of the computational set up in 2D
}
\label{scheme}
\end{center}
\end{minipage}
\end{figure}

The purpose of our study is to investigate the effects of the packing fraction of colloidal particles on various properties of the light 
incident on the system, including the path length, the amplification of the input light, the coefficients of transmission and reflection, the 
density distribution of scattering events along the axis orthogonal to the slab and the angular distribution of the output light. The 
geometric optics framework underlying the present study neglects wave interference and diffraction, as well as any non-linearity, due, for 
instance, to the depletion of excited states at high beam intensity. Our model, therefore, describes intensity feedback, which is also 
incoherent. On the other hand, the exact application of the geometric optics relations allows to match smoothly the ballistic regime of light 
propagation at low density of the scattering spheres with the diffusive regime at high density.
In a different but related context, the model and its results might be of interest in the analysis of reflectance and transmission of light 
from a turbid medium \cite{car, rogers, velle, turbid}.

\section{Model and method}
\label{method}
The present computational investigation has been inspired by the experimental study in Ref.~\cite{viola}, whose set-up consists of
$\sim 10^3$ reflective particle of macroscopic (mm) size immersed in an optically active fluid. The model and the experimental system, 
however, differ in two aspects. Since macroscopic heavy particles sediment under gravity, a fluid-like distribution has been obtained by 
applying mechanical shaking. Therefore, the experimental system is out of thermal equilibrium, with both the density of particles and their 
dynamics being controlled by shaking and gravity. These two effects are difficult to reproduce by simulation (at least because the details of 
shaking and the precise state of the solvent, i.e., density, temperature and viscosity, are not reported). Therefore, the simulation is 
carried out assuming thermodynamic (a-thermal, in the case of hard particles) equilibrium and neglecting gravity. In practice, the model is 
closer to the colloidal RLs than to the shaken granular system of Ref.~\cite{viola}. Moreover, the model assumes that the disordered 
configuration of the metal spheres is probed by a thin laser beam whose waist is significantly thinner that the particles diameter. Although 
the laser beam used in Ref.~\cite{viola} has a waist of a few mm, the thin beam assumption of the model greatly simplifies the analysis 
of the light path through the model system, which corresponds to the geometric scheme in Fig.~\ref{scheme}. Despite these two differences, we 
still emphasise the connection with the cited experimental paper, partly because, as in Ref.~\cite{viola}, our model assumes perfect 
reflection by sphere of diameter $\sigma >> \lambda$, with $\lambda$ being the wavelength of the input and output light. Moreover, the model 
system could be readily replicated by experiments with nearly the same set-up of Ref.~\cite{viola}, since laser beams of $\sim 50$ 
$\mu$m waist ($<< 2R$, see Fig.~\ref{scheme}) are widely available. 

More in detail, the simulation model consists of $N$ isotropic particles in 2D or 3D of coordinates $\{ {\bf R_i}, i=1, ..., N\}$ enclosed in 
an orthorhombic simulation box of size $L_x \times L_y (\times L_z)$, exposing two parallel surfaces perpendicular to the $x$ axis, and 
periodically repeated in the other direction(s). Confinement within the slab is enforced through an external potential $W(R^i_x)$ acting on 
the particles, which vanishes for $0 \leq R^i_x \leq L_x$ and is infinitely repulsive otherwise. In what follows, the term {\it disk} will be 
used for 2D particles only, while {\it sphere} will be used for 3D particles, but also when referring to either 2D and 3D systems without 
distinction.

For the sake of simplicity, we describe first the model and method for the 2D case, the extension to 3D being straightforward. The first
model that has been simulated consists of hard disks, whose hard-core (hc) potential energy is infinite if at least two disks overlap, and 
zero otherwise. The system is athermal, and the properties of the homogeneous phases and interfaces with confining hard walls (as in the 
present simulations) are well documented in the literature \cite{hd, interf}. For monodisperse homogeneous samples, all properties are 
functions of a single parameter i.e., the packing fraction $\eta=\frac{\pi}{4}\rho \sigma^2$, where $\rho$ is the number density. The 
homogeneous fluid phase is stable up to $\eta_F \sim 0.699$, separated from the hexagonal compact 2D crystal phase (stable for $\eta>0.723$) 
by the hexatic phase \cite{hd}. The highest (close) packing is $\eta_{cp}=\pi/\sqrt{12}=0.906899$. The amorphous phase might form by rapid 
compression of samples, but in monodisperse systems solidification into a defective crystal is the most likely result of increasing the 
density beyond $\eta_F$.

To cover also the liquid-vapour equilibrium, and especially the existence of a critical point, whose fluctuations might pose a particular 
challenge to simpler models of light propagation, simulations have also been carried out for 2D hard particles with an attractive potential 
well of depth $-\epsilon$ and amplitude $\delta$:
\begin{equation}
\label{fofr}
v(r)=\left\{
\begin{tabular}{ll}
$\ \infty$ \ \ \ \  & $\ \ \ r\leq \sigma$ \\
$-\epsilon$ \ & $\ \ \ \sigma < r \leq \sigma+\delta$ \\
\ 0  & $\ \ \ $ otherwise \\
\end{tabular}
\right.
\end{equation}
The particle-wall interaction $W(x)$ is still purely hard wall.

This model is no longer athermal, and we explored a few thermodynamic states in the ($\eta, T$) plane. The choice of these two parameters for
this second 2D system has been guided by the simulation results of Ref.~\cite{well}, considering, in particular, states close to the 
critical point of this model. Similar computations have been done for the 3D analogue of the force field model of Eq.~\ref{fofr}. In this 
second case, the critical point data of Ref.~\cite{krit3D} have been used.

In all simulated cases, both in 2D and in 3D, with and without the attractive potential well, particles are assumed to be fully reflective at 
the surface of their hard sphere potential.

For each system and choice of the thermodynamic conditions, $n$ independent configurations have been extracted from a MC simulation. For the
hard sphere plus square well system, the equilibration of the sample before selecting configurations has been verified by looking at the drift
of the potential energy, and the independence of the selected configurations has been assessed by estimating the potential energy 
autocorrelation time. For the pure hard disk model, the potential energy is constant and the validation has been carried out by computing the 
mean square displacement of particles as a function of MC time, making sure that, on average, particles explore a sufficient portion of space
to achieve equilibration and that the $n$ representative configurations are independent.

Each of the selected configurations has been used to generate $m >> 1$ paths for the propagation of light starting from the left of the slab, 
undergoing scattering events inside the slab, and exiting on the left (reflected) or on the right (transmitted) side of the sample (see 
Fig.~\ref{scheme} (a)). Here the assumption is that the motion of particles is so slow that the $m$ light paths are computed at fixed position
of the particles. Moreover, the beam of light is significantly thinner than the (macroscopic) particles' diameter, and the overall picture
corresponds to the scheme in Fig.~\ref{scheme}. Then, the beam scattering by the macroscopic reflecting particles is computed according to 
geometric optics.

First of all let us consider a single scattering event where a ray of incident radiation interacts with a scattering centre. Since the spheres
under consideration have a diameter $\sigma >> \lambda$ where $\lambda$ is the wavelength of the light, we can consider the scattering centres
as circular mirrors, and each interaction of a ray with a centre follows in accordance with specular reflection. 

A beam of light propagating through the medium is represented by a straight line, given in the parametric form:
\begin{equation}
x(t)=x_0+k_x (t-t_0)
\end{equation}
\begin{displaymath}
y(t)=y_0+k_y (t-t_0)\\
\end{displaymath}
where ${\bf k}\equiv (k_x, k_y)$ is a unit ($k_x^2+k_y^2=1$) propagation vector, and $t_0$ is an arbitrary starting parameter.
For simplicity, we will think of $t$ as time, and visualise light as traveling along the line at unit speed in the direction of increasing t.
In other terms, we imagine light moving as a thin beam, whose front is far less extended than the (macroscopic) size of the
particles.

In our analysis, the light beam starts at position $(x_0=-\sigma/2; y_0)$, where $y_0$ is a random number selected as specified in the 
Results section. The initial propagation vector is ${\bf k}={\bf k}_0\equiv (1; 0)$. During propagation in the random medium, traveling in the
generic $(k_x; k_y$) direction following a scattering event at position $(x_r; y_r)$, the time dependence of the beam location is given by:
\begin{equation}
x(t)=x_r+k_x t
\end{equation}
\begin{displaymath}
y(t)=y_r+k_y t
\end{displaymath}
where the origin of time has been shifted to the time of the previous scattering event. Elementary geometric considerations show that the 
distance between the centre $(R_x^i, R_y^i)$ of disk $i$ and the straight propagation
line is:
\begin{equation}
d_{min}=\sqrt{ d_0^2-\left[ k_x(R_x-x_r)+k_y(R_y-y_r)\right]^2}
\end{equation}
 where $d_0$ is the separation of $(x_r; y_r)$ from $(R_x^i, R_y^i)$, and occurs at:
\begin{equation}
t_{min}=k_x (x_r-R_x)+k_y (y_r-R_y)
\end{equation}
The beam intersects the surface of the same sphere provided $d_{min}\leq (\sigma/2)^2$, and the intersection (see Fig.~\ref{scheme} (b))
occurs at:
\begin{equation}
t_{s}=t_{min}-\sqrt{\frac{\sigma^2}{4}-d_{min}^2}
\end{equation}
The intersection is relevant only if $t_{s}\geq 0$. Then, the location of the following scattering event is identified as
$(x_r +k_x \bar{t}_s; y_r+k_y \bar{t}_s)$ where $\bar{t}_s$ is the minimum among the positive beam-surface intersection times with respect to
all particles.

The radius ${\bf R}$ joining the centre of the scattering sphere to the scattering point on its surface is orthogonal to the surface.
The incoming propagation vector ${\bf k}$  is projected into its components parallel (${\bf k}^{\parallel}$) and perpendicular
(${\bf k}^{\perp}$) to ${\bf R}$. The outgoing propagation vector ${\bf k'}$ is obtained by conserving the parallel component and reversing
the perpendicular one:
\begin{equation}
{\bf k'}={\bf k}^{\parallel}-{\bf k}^{\perp}
\end{equation}


Starting from the initial position $(x_0=-\sigma/2; y_0)$ and propagation vector ${\bf k}_0\equiv (1; 0)$ of the beam, scattering events are 
identified until there is no further scattering at positive time. If, at that point, the propagation vector has a positive $k_x$ component, 
the beam will exit the slab from its right interface, the path is counted as transmitted, and its length is accounted for in histograms of 
the properties of
transmitted paths. In this case, the number $n_s$ of scattering events since the beginning of the trajectory is also recorded to compute
the mean free paths of transmitted trajectories, as detailed below. In the case that $k_x < 0$, the beam exits the slab on the initial (left) 
side, and the path is counted among those reflected back.

This straightforward algorithm works and is easily adapted to incorporate periodic boundary conditions in the direction(s) orthogonal to 
$x$. However, at each reflection event, it requires to compute the particle-beam distance $d_{mim}$ of each particle in the system (or more, 
considering that the beam can cross a distance $>L_y$ in the transverse direction). 
A measure of time saving is introduced by considering that in 2D, starting from the scattering point $(x_r, y_r)$, a beam of wave vector
$(k_x > 0; k_y >0)$ can be scattered again only by particles whose position $(R_x^i; R_y^i)$ satisfies the conditions:
\begin{displaymath}
x_r -\frac{\sigma}{2} \leq R_x^i \leq L_x
\end{displaymath}
\begin{displaymath}
y_r-\frac{\sigma}{2} \leq R_y^i \leq y_r+k_y \left( \frac{L_x-x_r}{k_x}\right)
\end{displaymath}
with similar conditions for different sign combinations of the components $(k_x; k_y)$ of the scattering vector. These simple
geometric conditions decrease (on average by a factor of four in 2D, a factor of eight in 3D) the number of scattering events to be checked. 
Although this constant and relatively small gain factor is not decisive, for sample sizes of the order of $10^3$ particles the determination
of light paths is sufficiently fast to allow collecting statistics over several million paths for each sample, using only a few core hours
on a laptop or desktop.

All the equations reported in the previous paragraphs for 2D samples remain valid for the 3D case, with the trivial addition of the $z$ 
component in all coordinate and propagation vectors, playing a role completely analogous to that of the $y$ component. Also the considerations
on the relatively low computational cost remain valid in 3D.

The foremost characterisation of the transmission and amplification process concerns the length $l$ of the light paths crossing the slab after
many reflections by the colloidal particles, since this quantity is directly related to amplification. Because of the inherent disorder of the
distribution of scatterers, $l$ is a random variable whose probability distribution $P(l)$ will be characterised by simple statistical
parameters such as the mean value (Mean), the standard deviation (SDev), the skewness (Skew) and kurtosis, or, more precisely, the excess
kurtosis (ExKur), equal to the difference between the kurtosis and the constant value (=3) of the Gaussian distribution. All these parameters
are simply related to scaled moments (centred on the mean value and normalised by the SDev) of the distribution up to the fourth.
They are computed directly by the program that generates and analyses the light paths in the slab.

Besides the mean value, the analysis will focus on the excess kurtosis, which measures the {\it tailedness} of the distribution, with positive
values of ExKur (leptokurtic distribution) corresponding to an excess of positive outliers (i.e., $l>> \langle l \rangle$) with respect to the
Gaussian case. Because of the exponential character of light amplification by stimulated emission, these outliers might have a significant
impact on the intensity of outcoming rays, and represent the connection with the L{\'e}vy statistics aspects \cite{outliers, duf} of laser 
emission.

An additional important parameter to consider in multiple scattering is the mean free path, which is defined as the average distance travelled
by the ray between two scattering events \cite{principles}. For a single path of length $l$ resulting from $n_s$ scattering events, the
mean free path is:
\begin{displaymath}
l_s=\frac{l}{n_s}
\end{displaymath}
In the following section, the average mean free path along trajectories crossing the sample will be determined and discussed, defined as:
\begin{equation}
\langle l_s\rangle=\langle \frac{l}{n_s} \rangle
\end{equation}
where $\langle ... \rangle$ indicates the average over trajectories crossing the sample. A similar quantities could be computed for the paths 
being reflected.

A natural comparison for the $\langle l_s\rangle$ of simulation is the mean field result:
\begin{equation}
\langle l_s\rangle_{mf}=\frac{1}{\rho \sigma_s}
\label{uncorre}
\end{equation}
where $\sigma_s$ is the scattering cross section of the single scatterer, and $\rho$ is the density of scattering centres. A better 
semi-analytical expression for $\langle l_s\rangle$ accounting for lowest order effects of particle-particle correlations is available
\cite{sofk}. Using this approximation, however, already requires a simulation or an approximate theory to compute the static structure
factor of the particles, hence we find the simple, uncorrelated expression in Eq.~\ref{uncorre} to be the most suitable benchmark for the
computational results.

$\langle l_s\rangle$ is often used
as a diagnostic index to monitor the conditions for Anderson localisation \cite{and}, where it is scaled with the wave vector of the incident 
light \cite{ir}. However, the present investigation is purely geometric and the model does not display any non-linearity, hence Anderson 
localisation is beyond the scope of the present study. We can instead use the parameter $l_s$ as a single parameter characterisation of the 
diffusive medium through which the light travels. There are typically three regimes to describe multiple scattering: localisation, diffusive 
and ballistic \cite{principles}. In most cases, i.e., when the density is not too low, the propagation of light in the samples we considered 
occurs in the diffusive regime, where photons propagate according to the diffusion equation \cite{principles}. 

Another valuable parameter of the simulation is the amplification ratio of the output light to the input light. The present study, however,
is focused on the light propagation in the disordered medium filling the slab, while amplification is discussed only a posteriori, on the
basis of the results for the geometric trajectories. Wave-dependent aspects such as interference and diffraction are neglected, and also
neglected is the quantum nature of photon statistics and stimulated emission. In this simplified picture, each incoming ray carries the input 
intensity $i_0$. The output intensity $I_{out}$ of a ray going through the slab is estimated as:

\begin{equation} \label{eq:outputpower}
I_{out} = i_0 e^{\beta \langle l \rangle}
\end{equation}

where $\langle {l} \rangle$ is the average path length of transmitted rays and $\beta$ is the gain per unit length of the active fluid medium.
To estimate the order of magnitude of a few quantities, the value $\beta=0.002$ will be used, having clear that this value is arbitrary and
large, considering the unit of length adopted in this study, equal to the (presumably small) diameter $\sigma$ of particles. In reality, the 
value would depend on the physical parameters of the experiment, such as the chemical composition and concentration of the solution used and 
the size of the sample. If $N$ rays are sent through the slab then the input intensity is $N i_0$. On the other side of the slab we have 
$j << N$ rays with an energy of $i_0 e^{\beta \langle {l} \rangle}$. Therefore, the amplification ratio is:

\begin{equation} 
k_{ampl} = \frac{j}{N}e^{\beta \langle {l} \rangle}
\label{aaampl}
\end{equation}
These parameters and properties are discussed in more detail in Sec.~\ref{results}.

\section{Results}
\label{results}
 
\subsection{Colloidal particle configurations}

The first step in the simulation protocol has been to create random configurations of hard disks and spheres of unit diameter ($\sigma=1$) 
over a wide range of packing fractions. In the 2D case, for instance, as many non-overlapping disks as possible were packed into a square of 
side $L=40\sigma$. Using a random number generator, a sequence of $(x_i,y_i)$ centres were created, each time rejecting any disks overlapping 
with those already in the system. The maximum number of disks we could pack into the square was $N=1034$, corresponding to a packing fraction 
of $\eta = 0.537$. Then, a sequence of eleven samples spanning $0.01 \leq \eta \leq 0.5$ has been generated by expanding the box side to 
$L_x=L_y=\frac{\sigma}{2} \sqrt{N \pi/\eta}$. The list of packing fractions that have been simulated is: $\eta \in [0.01; 0.05\times j; \ 
j=1, ..., 10]$. A similar procedure in 3D results in a sequence of $11$ samples at the same $\eta(=\frac{\pi}{6} \rho \sigma^3)$ of the 2D 
case, but consisting of $5039$ spheres, a size somewhat larger but still comparable to that of the experimental sample in 
Ref.~\cite{viola}. Each sample in 2D and in 3D has been equilibrate by Monte Carlo for a number of steps ranging from $200 \times 10^6$ 
at $\eta=0.01$ to $10^9$ at $\eta=0.5$. A further MC run of $40 \times 10^6$ steps has been used to generate $201$ independent configurations 
per sample on which the analysis of light propagation has been carried out.

\begin{figure}[!htb]
\includegraphics[width=0.65\textwidth]{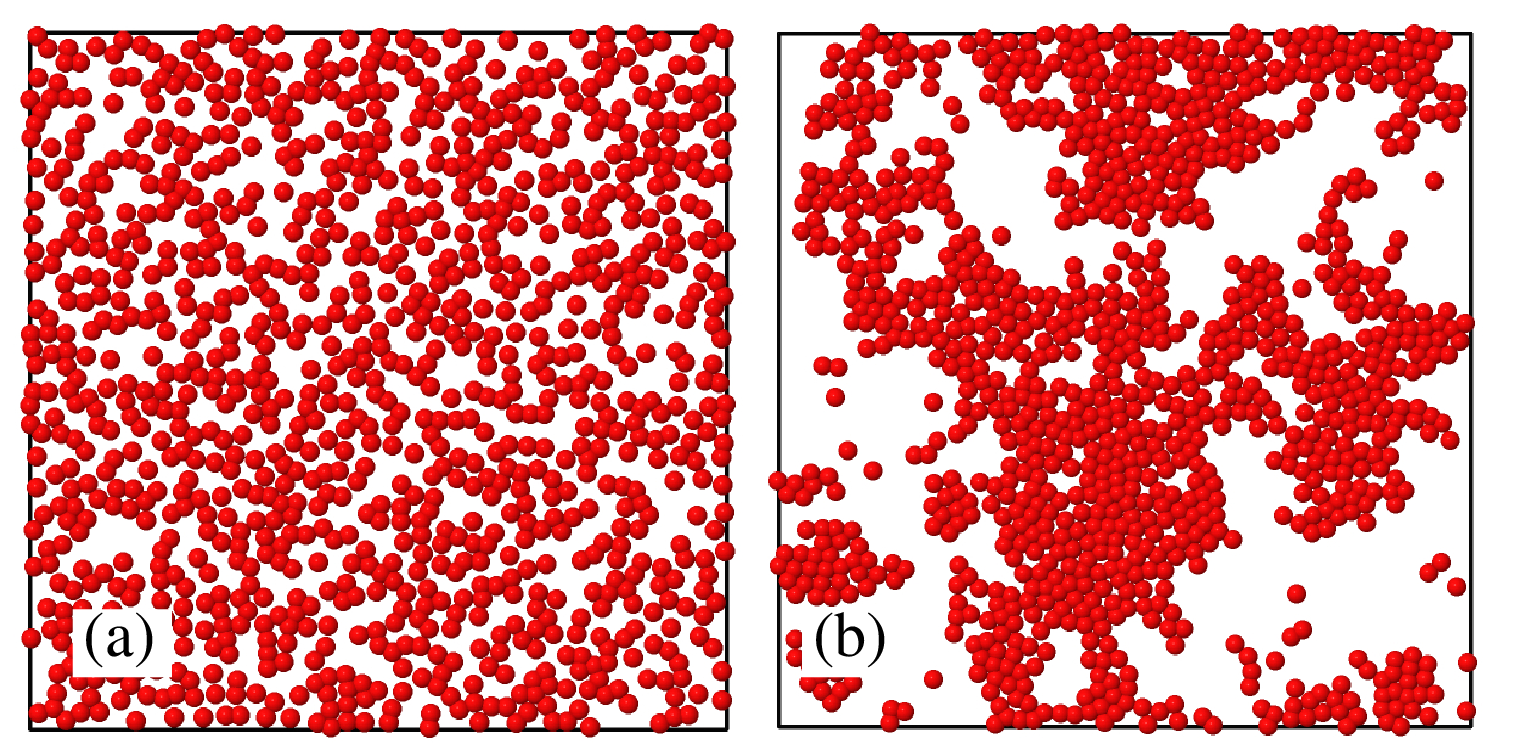}
\caption{\label{2DKrit} 
Snapshot of the simulation cell for two 2D systems: (a) the pure hard disk sample at $\eta=0.3$; (b) the hard disk plus attractive square well
potential in proximity of the critical point of the model at $\eta^{\ast}=0.2751$ and $T^{\ast}=0.5546\epsilon$. In (b), the strength of the 
attractive potential well is $\epsilon=1$ and the attractive square well extends over the interval $1 \leq r/\sigma \leq 1.5$. Both samples
consist of $1034$ particles, are finite along the horizontal $x$ direction, and periodic along the orthogonal $y$ direction. The radius of the
dots and the side of the cell are on scale.
}
\end{figure}

The starting configurations for the 2D and 3D hard particles are suitable to initiate simulations also for the system made of particles with
the square well attractive tail, since the two models share the same hard-core radius. Then, the equilibration and the generation of the $201$
independent configurations for the second model potential have been carried out in the same way as for the pure hc samples. Both in 2D and 3D,
samples have been considered for the attractive tail model, whose $(\rho; T)$'s belong to the critical isotherm for the two dimensionalities. 
The relevance of density fluctuations in proximity of the 2D critical state can be appreciated in Fig.~\ref{2DKrit}. The morphology in panel 
(a) and (b) of Fig.~\ref{2DKrit}, in particular, display the characteristic features of negatively (a) and positively (b) correlated random
structures \cite{davis}.

To isolate the effect of the slab surfaces, samples of the same size and number of particles have been generated by applying pbc in all 
directions. These samples, therefore, are homogeneous and lack both the density variation at the hard-wall surface, and the enhanced 
correlations parallel to the surface that have been discussed in a number of models. Also in this case, both 2D and 3D 
samples have been considered, with and without the attractive potential well, and simulated following the same protocol of the previous cases.
Then, the difference in the light propagation properties of homogeneous and finite-width samples is attributed to the presence of the 
hard-wall surfaces.

A final set of 2D samples has been generated adding a degree of polydispersivity to the population of particles. The polydispersivity
has been obtained by assigning the radius of each particle according to:
\begin{equation} 
\sigma_i=0.95 \ \sigma_0+0.1 \ \sigma_0 \zeta
\label{poly}
\end{equation}
where $\sigma_0=1$ and $\zeta$ is a random variable linearly distributed in $] 0; 1 [$. The size $L$ of each sample has been rescaled to
obtain the same $\eta$ values of the previous models.

\subsection{Analysis of light trajectories}

The preparation of configurations and the analysis of light paths arguably are simplest for the case of the the pure hard disks in 2D.
For this reason, we discuss first the results for this case, that represent a clear benchmark for all the other systems that have been
investigated. The analysis is based on the generation of at least one million light paths for each of the $201$ independent configurations 
($> 201 \times 10^6$ paths per sample) selected from the MC run of the 11 samples spanning the $0.01 \leq \eta \leq 0.5$ packing fraction 
range.

\begin{figure}[!htb]
\begin{minipage}[c]{\textwidth}
\vskip 0.7truecm
\begin{center}
\includegraphics[scale=0.60]{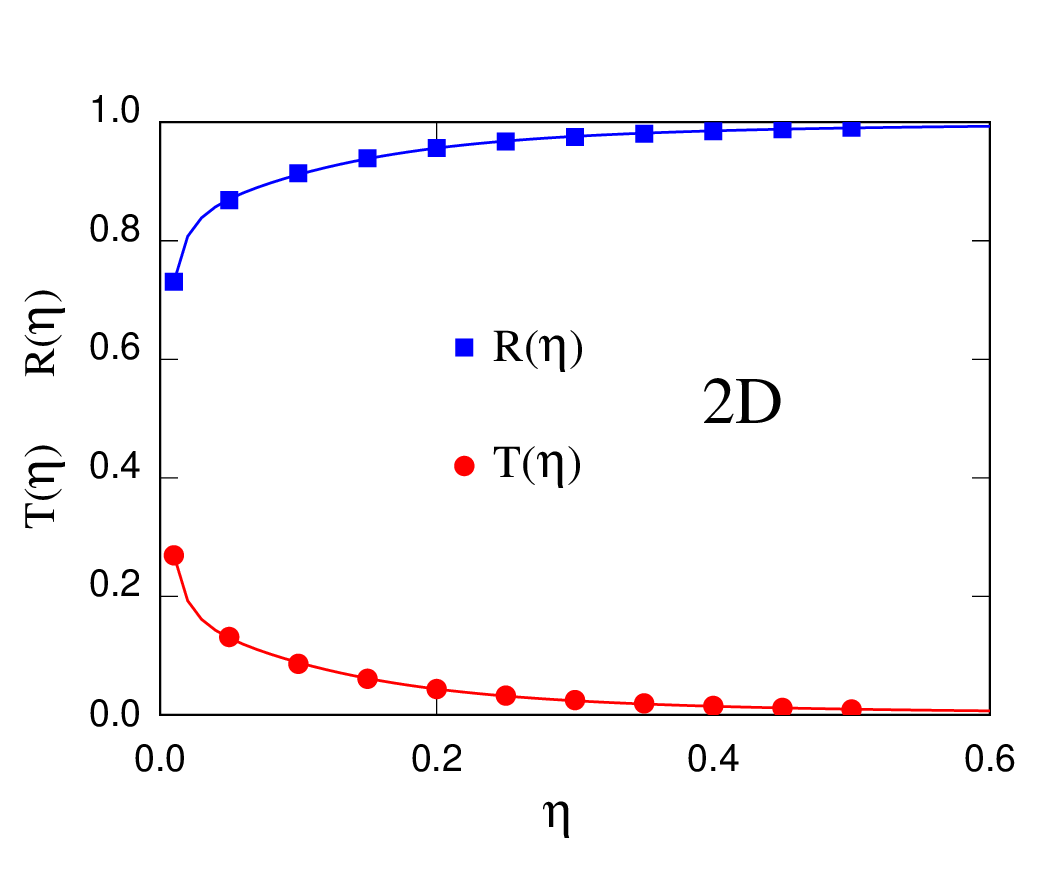}
\caption{Transmission ($T$, red dots) and reflection ($R$ blue squares) coefficients for the 2D slabs of purely repulsive hc particles as a 
function of packing fraction $\eta$. The full lines are the fit described by Eq.~\ref{pade}.
The number of particles is the same in all samples, therefore, the linear size of the slab decreasing with increasing $\eta$ like $L \propto
\eta^{-1/2}$
}
\label{rt}
\end{center}
\end{minipage}
\end{figure}

Preliminary to the discussion of all other properties, the transmission coefficient $T$ and the reflection coefficient $R$ are reported as
a concise characterisation of the overall effect of the slab on the incident beam. These coefficients are defined in purely geometric terms,
neglecting amplification by stimulated emission and absorption by the fluid medium. Then, $T$ is the fraction of light rays crossing the slab,
while $R$ is the fraction of light rays being reflected back to the original half space, with $T+R=1$. The results shown in Fig.~\ref{rt} 
follow intuition. At low $\eta$, $R$ is expected to grow linearly starting from its $\lim_{\eta \rightarrow 0} R(\eta)=0$, and to tend to $1$ 
with increasing $\eta$, without reaching it within the density range of fluid hard disk systems. The simple Pad{\'e} form:
\begin{equation}
R(\eta)=\frac{A\eta+B\eta^3}{1+C \eta+B\eta^3}
\label{pade}
\end{equation} 
provides an excellent fit of the data for $R$, and it can be trivially re-written for $T=1-R$. Assuming that the functional form
reflects the true dependence of $R$ on $\eta$, it is easy to verify that the deviation of $R$ from $1$ decreases like $\eta^{-2}$ with 
increasing $\eta$, and, equivalently, $T$ decreases like the same $\eta^{-2}$ with increasing $\eta$. Needless to say, considerations on 
asymptotic trends are not very relevant in this context, since the admissible range of $\eta$ is limited to $\eta\leq \eta_{cp}$. Apart from 
these general properties, the detailed values of $T$ and $R$ as a function of $\eta$ depend on the thickness of the slab. For the size of the 
simulated systems, it is apparent that the fraction of beams transmitted through the slab decreases very quickly at first, dropping below 
$10$\% already at $\eta=0.1$, then 
(necessarily) decreasing slowly with further increase of $\eta$.

\begin{figure}[!htb]
\begin{minipage}[c]{\textwidth}
\vskip 0.7truecm
\begin{center}
\includegraphics[scale=0.60]{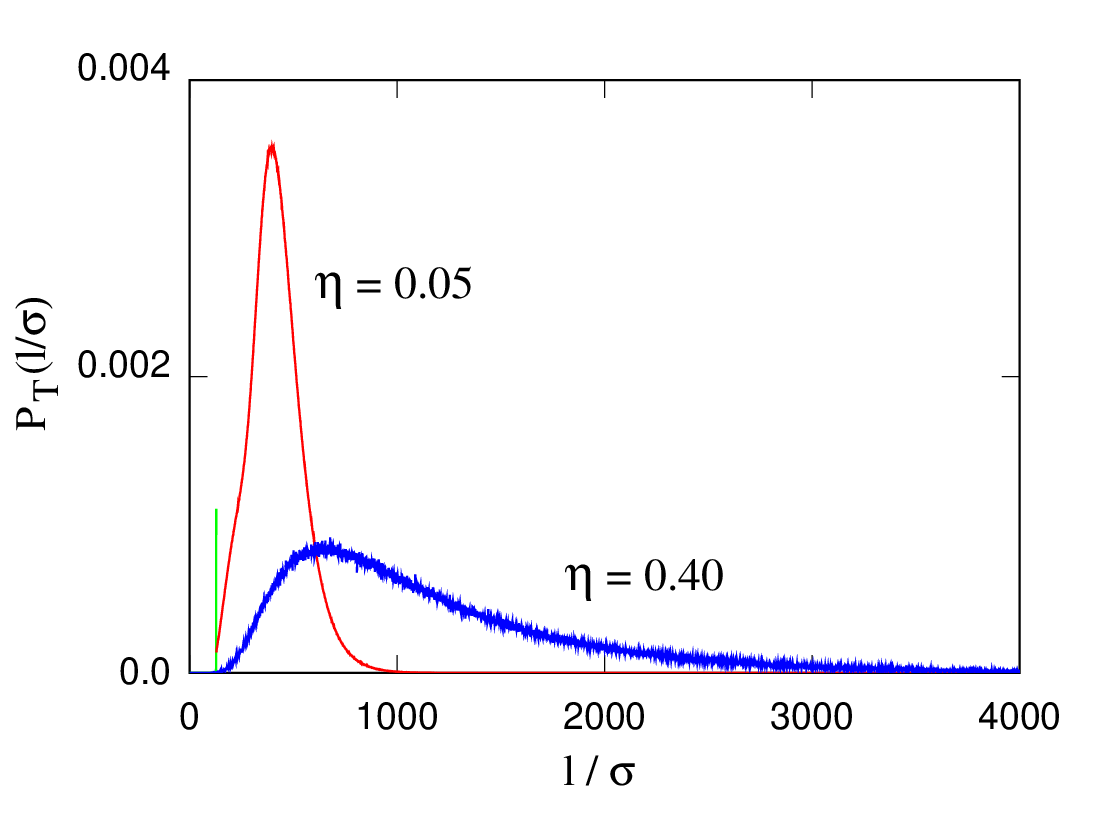}
\caption{Probability distribution $P_T(l)$ of the length $l$ of light path crossing the 2D slab. Particles interact with each other and
with the confining wall by purely hc interactions. The vertical green line belongs to the $\eta=0.05$ data, and represents the fraction of
light rays of length $l=L_x$ crossing the slab without being scattered.
}
\label{hist2D}
\end{center}
\end{minipage}
\end{figure}

The property that more closely reflects the scattering process of light in the slab is the length $l$ of light paths. As stated in 
Sec.~\ref{method}, this is a stochastic variable, characterised by the probability distribution $P(l)$. To be precise, the analysis is carried
out separately for transmitted and reflected paths, corresponding to two probability distributions $P_T(l)$ and $P_R(l)$, respectively.

The data for transmitted paths are discussed first. The generation of at least a few hundred million paths per sample allows 
to draw very accurate histograms of the paths' length, whose normalised version is the $P_T(l)$ shown in Fig.~\ref{hist2D} for $\eta=0.05$
and $\eta=0.4$.

\begin{figure}[!htb]
\begin{minipage}[c]{\textwidth}
\vskip 0.7truecm
\begin{center}
\includegraphics[scale=0.60]{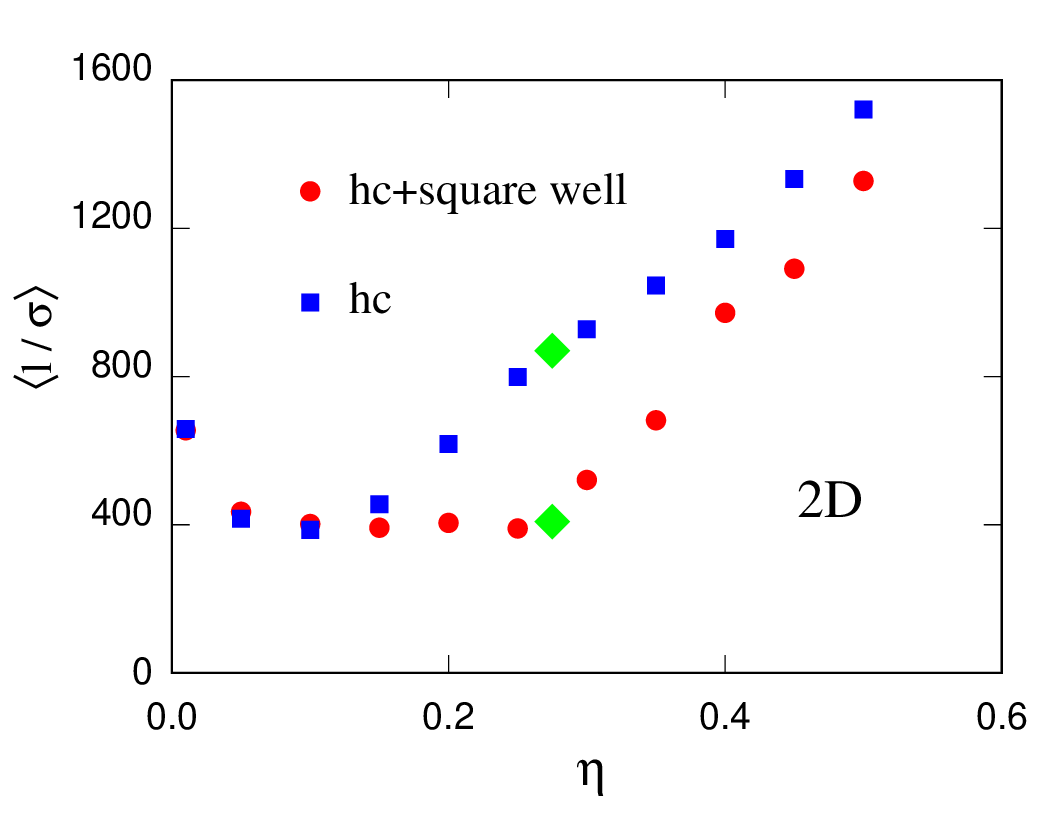}
\caption{Average length $\langle l \rangle$ of transmitted paths as a function of $\eta$. Blue filled squares: hc potential; red dots: hc plus
attractive square well tail. Green diamonds mark the values at the $\eta$ of the critical point for the 2D hs plus attractive square well
potential described in the text.
}
\label{al}
\end{center}
\end{minipage}
\end{figure}

The trends apparent in the figure are very intuitive. First of all, no transmitted path can be shorter than the width $L_x$ of the slab, hence
$P_T(l)=0$ if $l< L_x$. The sharp peak at $l=L_x$ in the plot for $\eta=0.05$, corresponds to rays crossing the slab without being scattered. 
This peak quickly becomes negligible with increasing $\eta$. To emphasise these features, the distribution $P_T(l)$ is rewritten as:

\begin{equation}
\label{smooth}
{P}_T(l)=\left\{
\begin{tabular}{ll}
$0$                 &  $l < L_x$ \\
$x\delta(l-L_x)$    &  $l=L_x$   \\
$(1-x)\bar{P}_T(l)$ &  $l>L_x$   \\
\end{tabular}
\right.
\end{equation}
where $0 \leq x \leq 1$ is the relative weight of the $\delta$-like peak. The continuous portion $\bar{P}_T(l)$ of the distribution, whose 
normalisation is one, is skewed right and, with increasing $\eta$, it becomes broader and moves toward larger $l$'s, despite the progressive 
shrinking of the slab. 

\begin{table}
 \begin{tabular}{||c| c| c| c| c| c| c| c| c| c| c| c||}
  \hline
 \multicolumn{12}{|c|}{Transmitted Path Length Distribution: Statistical Moments} \\
 \hline
 \textbf{$\eta$} & \textbf{0.01} & \textbf{0.05} & \textbf{0.1} & \textbf{0.15} & \textbf{0.2} & \textbf{0.25} & \textbf{0.3} & \textbf{0.35} & \textbf{0.4} & \textbf{0.45} & \textbf{0.5}\\ [0.5ex] 
 \hline\hline
 \textbf{Mean } & 658 & 416      & 386   & 455   & 618   & 798   & 928     & 1046   & 1172   & 1333   & 1521    \\ 
                & (-) & (452)    & (391) & (454) & (639) & (840) & (963)   & (1076) & (1200) & (1360) & (1548)  \\ 
 \hline
 \textbf{SDev} & 302 & 128       & 140   & 223   & 370   & 507   & 590     & 665    & 745    & 851    & 969     \\
                & (-) & (139)    & (135) & (219) & (412) & (582) & (654)   & (724)  & (802)  & (904)  & (1027)  \\ 
 \hline
\textbf{Skew} & 0.686 & 0.574    & 1.36  & 1.70  & 1.77  & 1.79  & 1.81     & 1.80   & 1.79   & 1.81   & 1.80   \\
                & (-) & (1.29)   & (1.35)& (1.73)& (2.15)& (2.23)& (2.15)   & (2.11) & (2.08) & (2.06) & (2.04) \\ 
\hline
 \textbf{ExcKur} & 0.1 & 1.07    & 3.40  &  4.7  & 4.9   & 5.1   & 5.2      & 5.1    & 5.0    & 5.2    & 5.1    \\
                & (-) & (2.77)   & (3.02)& (5.00)& (7.71)& (8.29)& (7.72)   & (7.43)  & (7.22) & (7.07) & (6.96) \\ 
 \hline
 \hline
\end{tabular}
\caption{Mean value (Mean), standard deviation (SDev), skewness (Skew) and excess kurtosis (ExKur) of the probability distribution $P_T(l)$ 
for the length $l$ of transmitted paths in 2D samples of hard disks as a function of the packing fraction $\eta$. The corresponding values
obtained by the fit of the continuous part $\bar{P}_T(l)$ of $P_T(l)$ with an inverse Gaussian distribution are listed in parentheses. The 
fit parameter have been omitted for $\eta=0.01$ because in that case the fit is not even qualitatively close to the simulation data.
Statistical error bars are implicitly indicate by the number of digits of the data.
}
\label{tabI}
\end{table}

The qualitative results obtained by visual inspection of the $P_T(l)$ distributions are quantified by computing the average path length 
$\langle l\rangle$, the variance, skewness and kurtosis of the length distribution of all samples. The parameters collected in Tab.~\ref{tabI}
show that for $\eta\geq 0.1$ the mean value and the standard deviation increase rapidly with increasing $\eta$, while both skewness and excess
kurtosis saturate to a constant value from below. The excess kurtosis, which is of particular interest for the present discussion, is 
always positive and sizeable, implying that the distribution has an excess of outliers with respect to the Gaussian distribution.
To be precise, the average length has a non monotonic behaviour, at first decreasing in a narrow interval of 
$\eta$, with a minimum at $\eta\sim 0.1$ (see Fig.~\ref{al}).
The presence of the minimum has to be expected since, at constant number of particles, the linear size of the system, and therefore the 
minimum path length, diverge in the $\eta\rightarrow 0$ limit, while at large $\eta$ the value of $l$ increases again because of trapping of
light into the slab. 

While the $\delta$-like peak is relevant only at low $\eta$ and its weight $x$ is difficult to predict, a few more observations can be made on
the continuous part $\bar{P}_T(l)$ of the distribution. First, one can remark that as soon as the density of scattering centres is 
non-negligible, the light paths inside the slab resembles the trajectory of a random walk. The length $l$ of each of the transmitted paths is 
equivalent to the time $t$ spent by a Brownian particle to travel at unit velocity from $(-\sigma/2; y_0)$ to the exit point having $x=L_x$ 
(irrespective of the $y$ exit coordinate). The distribution of these times has a vague relation with the Wald distribution $P_W$, also known 
as the inverse Gaussian distribution:
\begin{equation}
P_W(t\equiv l; \mu, \lambda)=\sqrt{\frac{\lambda}{2\pi t^3}} \exp{\left[ -\frac{\lambda (t-\mu)^2}{2\mu^2 t}\right]}
\label{wald}
\end{equation}
which describes the distribution of times taken by a Brownian particle to travel a given distance. In Eq.~\ref{wald}, $\mu$ is the average
value and $\lambda$ is a fit parameter closely related to the variance of the distribution. Admittedly, because of the spread of $y$-exit 
coordinate along the surface plane, the random variable $l$ measured in the simulation is not the time needed to a Brownian particle to drift 
by the distance $L_x$, therefore, Eq.~\ref{wald} cannot be the a priori model that describe the simulation results. Nevertheless, we verified 
that Eq.~\ref{wald} still represents a suitable analytical form to fit the simulation results, provided the density of scattering centres is 
not too low. 

To be precise, the expression of Eq.~\ref{wald} cannot reproduce the simulation data because, as already remarked, no transmitted path
can be shorter than the width $L$ of the slab, and $P(l)$ has to be zero for $l< L$.

To enforce this condition, the smooth part of the probability distribution is modelled as:
\begin{equation}
\label{newald}
\bar{P}_T(l)=\left\{
\begin{tabular}{ll}
$0$                 &  $l < L_x$ \\
$P_{W}(l-L_x)$ &  $l \geq L_x$   \\
\end{tabular}
\right.
\end{equation}
where $P_W$ is the Wald distribution of Eq.~\ref{wald}. The rigid shift by $L_x$ changes the average $\langle l \rangle$ from $\mu$ to 
$\mu+L_x$ but it does not change the estimate of the variance $\mu^3/\lambda$, skewness $3\sqrt{\mu / \lambda}$ and excess kurtosis 
$15 \mu /\lambda$. 

\begin{figure}[!htb]
\begin{minipage}[c]{\textwidth}
\vskip 0.7truecm
\begin{center}
\includegraphics[scale=0.60]{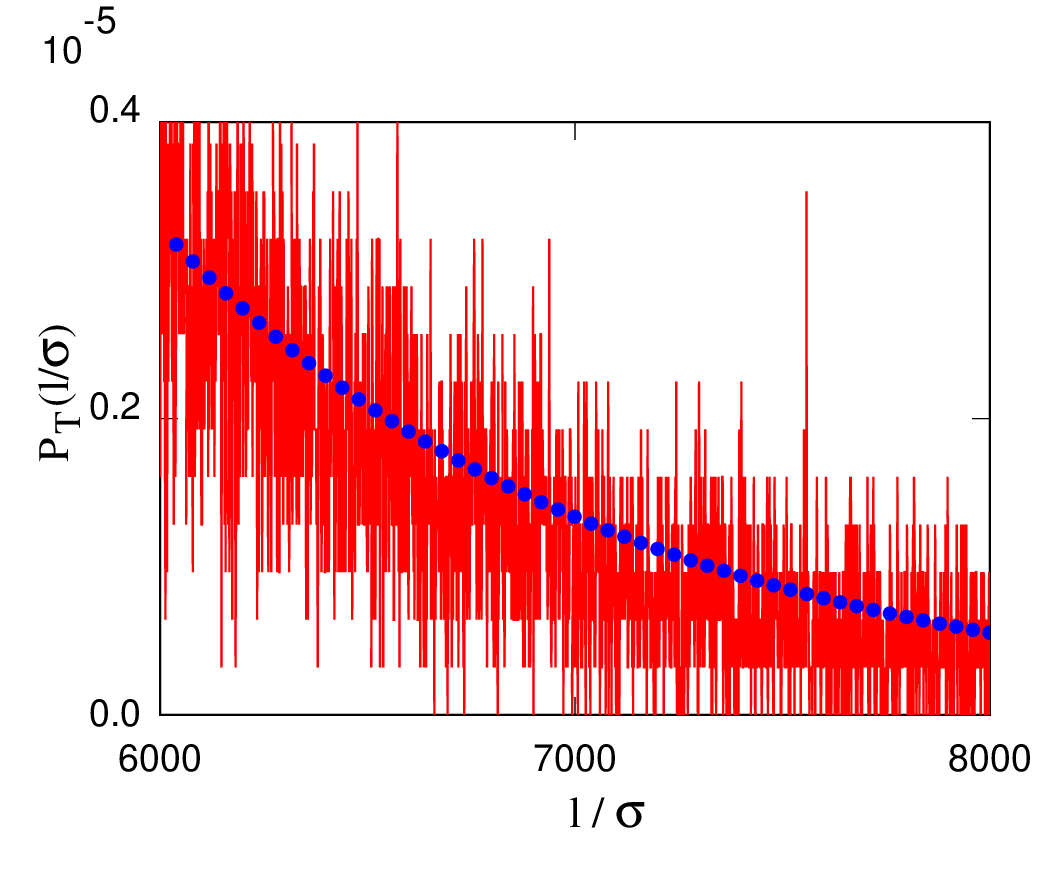}
\caption{Long-$l$ tail of the probability distribution $P(l)$ of light path crossing the 2D slab at $\eta=0.5$. 
Comparison of fit (blue dots) and raw data (red line).
}
\label{compaa}
\end{center}
\end{minipage}
\end{figure}

As soon as $\eta$ exceed $\sim 0.1$, the inverse Gaussian distribution of Eq.~\ref{newald} provides an excellent fit of the entire 
distribution given by simulation (see Fig.~S1 in the Supplemental Material \cite{supple}), allowing an immediate estimate of average, 
standard deviation, skewness and excess kurtosis, whose values are close to those computed directly by the simulation (see Tab.~\ref{tabI}). 
In the medium/high density range, the 
ability of the inverse Gaussian distribution to fit the simulation data extends to high $l$, as shown in Fig.~\ref{compaa}. This observation 
is remarkable, since the tail of the distribution has a negligible weight in the fit. Perhaps more importantly, the analytical expression of
the fit distribution allows to derive the asymptotic distribution of the output power, following amplification by the optically active medium.
This aspect is discussed in Sec.~\ref{ampll}.

\begin{figure}[!htb]
\begin{minipage}[c]{\textwidth}
\vskip 0.7truecm
\begin{center}
\includegraphics[scale=0.60]{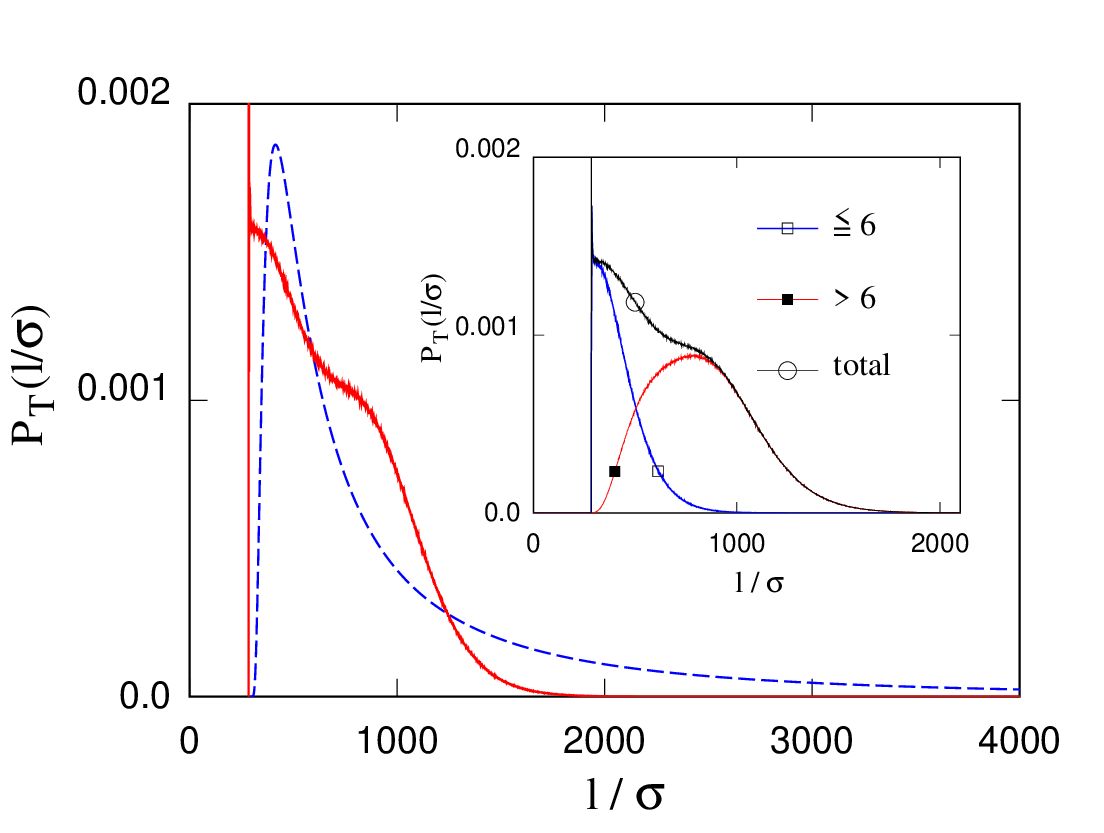}
\caption{Main panel: Probability distribution $P(l)$ of the length $l$ of light path crossing the 2D slab at $\eta=0.01$. Comparison of 
fit (full line) and simulation data (dash line). The inset shows the breakdown of the simulation data into contribution from paths undergoing 
$\leq 6$ and $>6$ scattering events while crossing the slab. The different curves are identified by the symbols on the lines, as specified 
in the figure.
}
\label{compab}
\end{center}
\end{minipage}
\end{figure}

It is apparent, however, that at low $\eta$ the inverse Gaussian 
distribution differs even qualitatively from the simulation data, as shown in Fig.~\ref{compab} for the $\eta=0.01$ case. 
The reason of the discrepancy between simulation data and inverse Gaussian fit at low $\eta$ is easy to guess and verify through 
the data analysis. The usage of the inverse Gaussian distribution relies on the underlying diffusive dynamics for the propagation of light in 
the slab, which, in turn, implies a multitude of scattering events to approach a genuine Brownian process. At low $\eta$, instead, a sizeable
number of paths cross the slab without being scattered, and many other paths are scattered only a limited number of times, thus the limit of a
Brownian process is not reached. This interpretation is verified in the inset of Fig.~\ref{compab} (b), showing the contribution to $P(l)$ 
from paths that cross the slab undergoing $\leq 6$ and $>6$ scattering events. The two sets of paths seem to represent different populations, 
accounting for the bimodal character of the distribution, that cannot be reproduced by the inverse Gaussian analytical expression.

Turning now to the properties of the reflected rays: for all 2D hard disk samples, the probability distribution $P_R(l)$ for the length $l$ 
of the reflected rays decays monotonically as a function of $l$. Moreover, the average path length $\langle l \rangle_R$ decreases 
monotonically with increasing $\eta$ (see Tab.~\ref{tabR2D}). The decrease is very rapid at first, then the value nearly saturates for 
$\eta \geq 0.30$, apparently because of the compensation between the increasing reflectivity of the slab, that reduces $\langle l \rangle_R$, 
and the increasing length of any path that is able to penetrate beyond the first layer of colloidal particles before being reflected back. 

A logarithmic plot of $P_R(l)$ suggests the functional form:
\begin{equation}
P_R(l)=\frac{A}{1+B l^{1/2}+C l^{3/2}}
\end{equation}
as resembling the simulation data at all values of $\eta$. The numerical fit is reasonably accurate, but admittedly the fit is empirical
and not as good as the inverse Gaussian distribution for $P_T(l)$. On the other hand, a fit with a sum of exponentials does not achieve the
same $\chi^2$ (accounting for the number of degrees of freedom in the fit). It might be useful to remark that $P_T(l)$ does not measure,
nor it is directly related to the penetration depth of the rays into the slab. 

\begin{table}
 \begin{tabular}{||c| c| c| c| c| c| c| c| c| c| c| c||}
  \hline
 \multicolumn{12}{|c|}{Reflected Path Length Distribution: Statistical Moments} \\
 \hline
 \textbf{$\eta$} & \textbf{0.01} & \textbf{0.05} & \textbf{0.1} & \textbf{0.15} & \textbf{0.2} & \textbf{0.25} & \textbf{0.3} & \textbf{0.35} & \textbf{0.4} & \textbf{0.45} & \textbf{0.5}\\ [0.5ex] 
 \hline\hline
 \textbf{Mean } & 320 & 120      &  80   &  70   & 67   &  66   &  62     &   59   & 57     &  57    &  56     \\ 
 \hline
 \textbf{SDev} & 330 & 150       & 125   & 140   & 170   & 200   & 210     & 220    & 225    & 240    & 255     \\
 \hline
\end{tabular}
\caption{Mean value (Mean) and standard deviation (SDev) of the probability distribution $P_R(l)$ for the length $l$ of reflected paths in 2D 
samples of hard disks as a function of the packing fraction $\eta$. Statistical error bars are implicitly indicate by the number of digits of the data.
}
\label{tabR2D}
\end{table}

The mean-free path $\langle l_s\rangle$ defined as the average length of linear segments in between scattering events, 
decreases monotonically with increasing $\eta$. A fit with a simple inverse power of $\eta$ ($\langle l_s\rangle=A/\eta$), following the mean 
field result, reproduces the simulation data fairly well. A significant further reduction of the square deviation and of $\chi^2$ is obtained 
by adding a logarithmic term, i.e.: $\langle l_s\rangle=A/\eta+B/\log{(\eta)}$, that apparently arises from effects beyond mean field. The 
monotonic decrease of $\langle l_s\rangle$ and the non-monotonicity of $\langle l \rangle$ versus $\eta$ can be matched by considering that 
the number of scattering events experienced by a beam transmitted through the slab increases slowly at first, and then more rapidly with 
increasing $\eta$, as verified by the simulation data. To a very good approximations, these properties of the mean-free path are shared by 
transmitted and reflected trajectories.

\begin{figure}[!htb]
\begin{minipage}[c]{\textwidth}
\vskip 0.7truecm
\begin{center}
\includegraphics[scale=0.60]{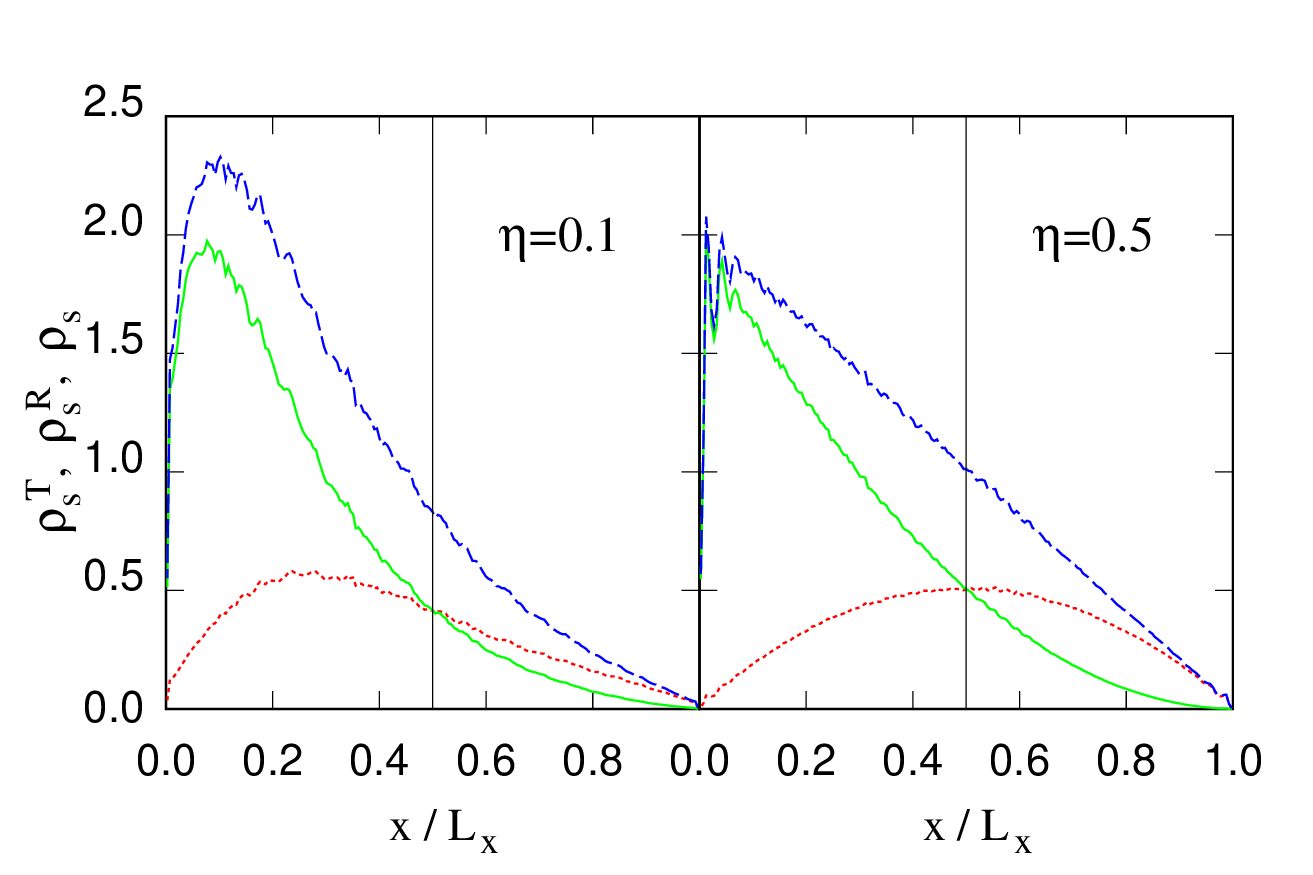}
\caption{Probability density of scattering events across a 2D slab whose particles interact via a purely hc potential. 
Red dotted line: $\rho_s^T(x)$; green line: $\rho_s^R(x)$; blue dash line: $\rho_s(x)=\rho_s^T(x)+\rho_s^R(x)$. 
}
\label{rhos}
\end{center}
\end{minipage}
\end{figure}

Given the inhomogeneous and asymmetric nature of the system and process, it is of interest to determine where the scattering events take place
along the coordinate $x$ normal to the slab. This information is contained in the density distribution $\rho_s(x)$ of scattering events, shown
in Fig.~\ref{rhos} (a) and (b), at low and high $\eta$, respectively. The properties of the system are easier to understand at high $\eta$ 
such that the propagation of light is diffusive. In such a case, the propagation of light is similar to many other transport processes based 
on diffusion with superimposed a slight drift, such as electric and thermal conductivity, driven by an electrostatic potential and temperature
difference, respectively. In the present case, the potential is represented by the light intensity $I(x, t)$ (where $t$ is time), the 
resistance is due to the scattering processes, that limit the flow of the light current $J(x,t)=k\ [d I(x,t)/ dx]$ which, because of light 
intensity conservation, satisfies a continuity equation. In the previous relation, $k$ is the light conductivity. At stationary conditions, 
$J(x, t)$ is constant along $x$ and $t$, reflecting the constant gradient of the light intensity within the slab along $x$. In other terms, 
at stationary conditions, the light intensity $I(x)$ decreases linearly with $x$. Neglecting possible deviations due to the layering of 
particles along $x$ in proximity of the hard walls, the density $\rho_s(x)$ of scattering events is proportional to the light intensity 
$I(x)$, hence the probability distribution $\rho_s(x)$ of scattering events (considering both transmitted and reflected paths) has to decrease
linearly from left to right across the slab. This qualitative picture is born out by the simulation results at high $\eta$, reported in 
Fig.~\ref{rhos} (b), showing a clear linear dependence of $\rho_s(x)$ on $x$. The picture, however, is violated at low $\eta$ (see 
Fig.~\ref{rhos} (a)) since in that case, the propagation of light is at least partly ballistic, and the influence of the interfaces extends 
to a sizeable portion of the whole system.

More telling is the partitioning of $\rho_s(x)$ into the contributions $\rho_s^T(x)$ and $\rho_s^R(x)$, counting the number of scattering 
events arising from transmitted and reflected paths, respectively. In the diffusive regime, paths at the geometric centre of the slab have 
nearly the same probability of exiting on the left and on the right of the slab. Therefore, to a good approximation, $\rho_s^T(L_x/2)=
\rho_s^R(L_x/2)$, provided the propagation is diffusive. A slight unbalance might be due to the random position of the colloidal particles, 
and especially to the weak constant flow $J$ crossing the slab from left to right. 

Furthermore, $\rho_s^T(x)$ has to be nearly symmetric around $x=L_x/2$, with $d \rho_s^T(x)/ dx=0$, since transmitted paths can be traversed 
in both directions, hence the population of transmitted paths is the same in the right and left directions, and the same distribution of 
scattering events is shared by the propagation from left to right and vice-versa. The only distinction between the left and right of 
$\rho_s^T(x)$ is in the boundary conditions and in the net flow $J$, that do not affect the details of the scattering distribution far from 
the surface walls, always provided $\eta$ is high and the light propagation diffusive. At low $\eta$, the symmetry-breaking current $J$ is 
high, the effect of the walls is less screened close to the surfaces, and the asymmetry in $\rho_s^T(x)$ is sizeable, as shown in 
Fig.~\ref{rhos} (a). Besides these qualitative features, the quantitative details of $\rho_s^T(x)$ and $\rho_s^R(x)$ are determined by the 
strength of the scattering processes, which depends on $\eta$, and by the boundary conditions, including the fact that $\rho_s^R(x)$ has to 
vanish at $x=L_x$.

\begin{figure}[!htb]
\begin{minipage}[c]{\textwidth}
\vskip 0.7truecm
\begin{center}
\includegraphics[scale=0.60]{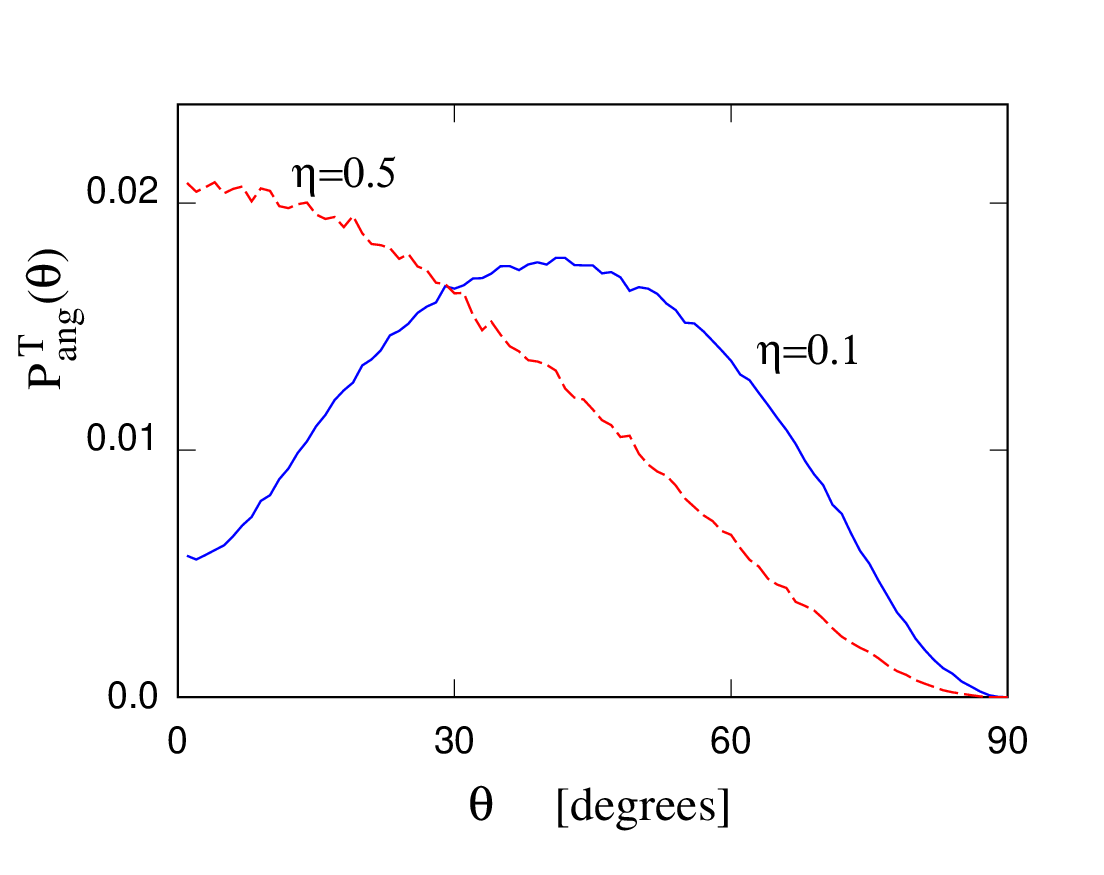}
\caption{Probability distribution of the exit angle $\theta$ of transmitted rays through 2D samples measured with respect to 
$\hat{\bf x}$. Blue full line: $\eta=0.1$; red dash line: $\eta=0.5$.
}
\label{direct}
\end{center}
\end{minipage}
\end{figure}

The direction of the outcoming rays is an important aspect in view of applications of the system as random laser. For transmitted rays, the 
simulation results show that at low $\eta$ the probability distribution $P^T_{ang}(\theta)$ of angles $\theta$ with respect to the 
$\hat{\bf x}$ direction presents a broad peak around $\theta=45^{\circ}$, and a non-negligible value in the forward direction, 
$\theta=0^{\circ}$ (see Fig.~\ref{direct}). At $\eta> 0.4$, the $P^T_{ang}(\theta)$ distribution is peaked at $\theta=0^{\circ}$, but still 
broad. It has been verified that path lengths $l$ and exit direction $\theta$ are uncorrelated. Directional beams, therefore, can only be 
obtain by collimation, sacrificing most of the intensity. For reflected rays, the probability distribution $P^R_{ang}(\theta)$ of the angle 
$\theta$ is as broad as for the transmitted ones, but, especially at low $\eta$, it peaks at $180^{\circ}$, possibly because their direction 
is less randomised before leaving the slab. This is supported by the fact that at high $\eta$, when the beam undergoes a higher number of 
scattering events in the skin-depth subsurface layer before being back-reflected, the $180^{\circ}$ peak is less pronounced than at low 
$\eta$ (see Fig.~S2 in Ref.~\cite{supple}).

Analysis of the simulation trajectories shows that the propagation direction does not loose its memory at every step, as assumed by the
simplest random walk models, nor it is equal to the average scattering angle from the single isolated scatterer. In the simulations, the 
autocorrelation function of the propagation unit vector ${\bf k}$ shows a limited memory, lasting a few scattering events. At densities 
$\eta \geq 0.01$, in particular, the average $\langle {\bf k} \cdot {\bf k'}\rangle$ for propagation vectors separated by one scattering 
event is negative both for transmitted and reflected rays. In other terms, the short time correlation function of propagation vectors is 
dominated by backscattering both along transmitted and reflected trajectories. This feature has to change at very low density, since in the 
limit of a single scattering sphere the same autocorrelation function can be computed analytically and (obviously) it is positive for 
transmitted paths, and negative for reflected paths, as we also verified numerically going to very low $\eta$. The same 
$\langle {\bf k} \cdot {\bf k'}\rangle$ resulting from each scattering event in a colloidal solution is discussed experimentally in
Ref.~\cite{corre}.

\begin{figure}[!htb]
\begin{minipage}[c]{\textwidth}
\vskip 0.7truecm
\begin{center}
\includegraphics[scale=0.60]{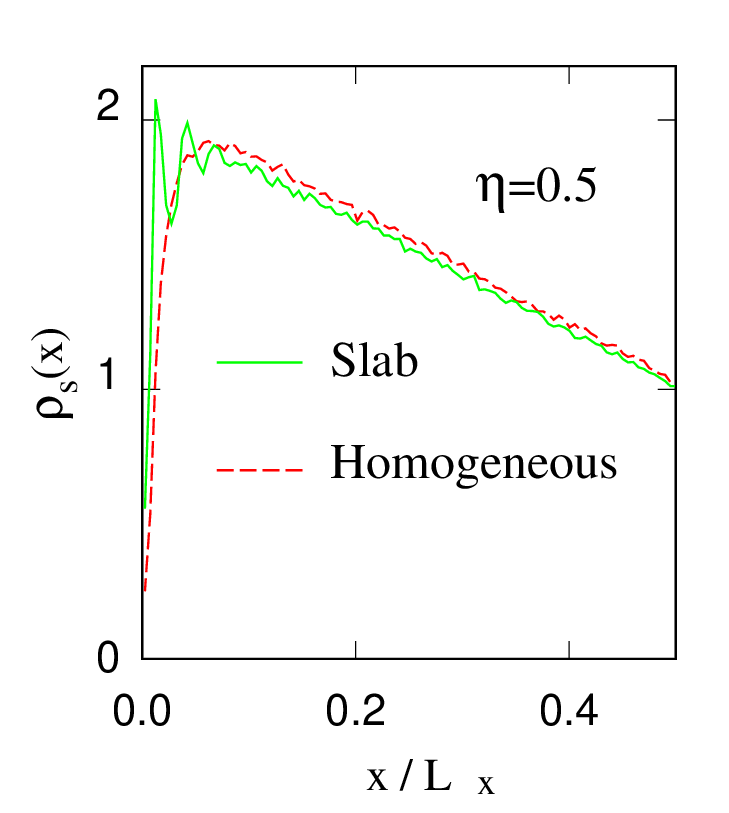}
\caption{Probability distribution of scattering events across the slab (green full line) and the homogeneous sample (red dash line) for 
$\eta=0.5$. The $\rho_s(x)$ reported in the figure refers to the total number of scattering events, belonging to both reflected and 
transmitted light paths. Most of the change observed for $x/L_x \leq 0.1$ concerns the reflected paths.
}
\label{Rpbc}
\end{center}
\end{minipage}
\end{figure}

One of the remarkable aspects of hc fluids confined by hard walls is the density pile up at the interface (see Fig.~S3 in 
Ref.~\cite{supple}), due to the requirement of mechanical stability ($p_{xx}(x)=const$, where $p_{xx}$ is the $xx$ element of the stress
tensor) across the slab \cite{mech}.
In these systems, the peak density at contact increases rapidly with increasing packing fraction, and at high $\eta$ it might severely hamper 
the penetration of light into the slab, increasing its reflectivity.  To quantify this effect, comparison has been made with the results 
computed for a sample of the same size and packing, but rendered homogeneous by the application of pbc in all directions. It turns out that 
the overall effect, measured for instance by the transmission and reflection coefficients, or by the average length of transmitted and 
reflected path, is small at all $\eta$, partly because at the high $\eta$ at which the density pile up is most important, the transmission of 
light through the slab is already effectively suppressed by the high average particle density. The effect on the microscopic scattering 
mechanisms, however, is important, and easily identified by the computation. The comparison of the distribution of scattering events, for 
instance, shown in Fig.~\ref{Rpbc}, shows an important change in the fine structure of $\rho_s(x)$ in proximity of the interface. The value of
$\rho_s(x)$ is a measure of the light intensity at position $x$. The slightly lower value of $\rho_s(x)$ well inside the sample (see again 
Fig.~\ref{Rpbc}) is due to the higher reflectivity of the inhomogeneous slab, although the effect integrated over the slab width $L_x$ is 
only minor.

The last remark on hc 2D systems concerns the effect of polydispersivity, that has been modeled as specified by Eq.~\ref{poly}. Also in this
case, the effect grows slowly with increasing $\eta$ but it is small at all $\eta$'s that have been simulated. At $\eta=0.5$, for instance,
both the transmission coefficient and the average path length of the polydisperse sample are about $1$ \% higher than those of the 
monodisperse sample. The two trends are somewhat contradictory, since an increase of path length points to a lower transmission of light.
However, the difference in both $T$ and $\langle l \rangle$ is beyond the error bar. Perhaps more telling, the differences between mono- and 
polydisperse samples are systematic being similar over the entire $\eta$ range, confirming that they are a genuine effect of polydispersivity.
We verified that especially at high density, there is a relatively weak preferential segregation of small particles at the interface, changing
the reflection/transmission properties of the layered slice of the slab in contact with the hard wall, and possibly causing the changes in 
all the other properties.

\subsection{Hard disks with an attractive square well tail}

Monte Carlo simulations and the analysis of light trajectories have been carried out for 2D particles interacting with a hard disk potential 
plus an attractive tail, following the same protocol used in the case of purely hard disks. Although the two models are fairly similar, there 
are at least two characteristic differences. At variance from pure hard disk case, the phase diagram of the hard disk plus attractive tail 
potential usually (but not always, see Ref.~\cite{toad1, toad2}) has a genuine liquid phase and a critical point whose critical 
fluctuations might greatly affect light paths in the slab, since opalescence is precisely one manifestation of criticality. The second 
difference concerns the structure of the fluid at the hard wall surface. Assuming that the wall-particle potential $W(x)$ is purely hard-core,
the contact density of particles $\rho(x=0)=\rho(x=L_x/2)$ depends directly on the constant value of the stress tensor element $p_{xx}(x) >0$ 
across the slab in the direction orthogonal to its surface. In the purely hard disk case, pressure tends to be high, and the interface is an 
accumulation point for particles. Density, therefore, has a peak at the interface, and decays towards the bulk value with wide amplitude 
oscillations. For particles interacting with the hc plus attractive tail, the value of $p_{xx}$ tends to be moderate up to fairly high 
$\eta$, the interface is depleted of particles, and the density does not display short-wavelength oscillations at contact (see Fig.~S3 in 
Ref.~\cite{supple}). These differences in the liquid structure, although localised at the interfaces, might affect the transmission and 
reflection coefficients of the whole slab.

\begin{figure}[!htb]
\begin{minipage}[c]{\textwidth}
\vskip 0.7truecm
\begin{center}
\includegraphics[scale=0.60]{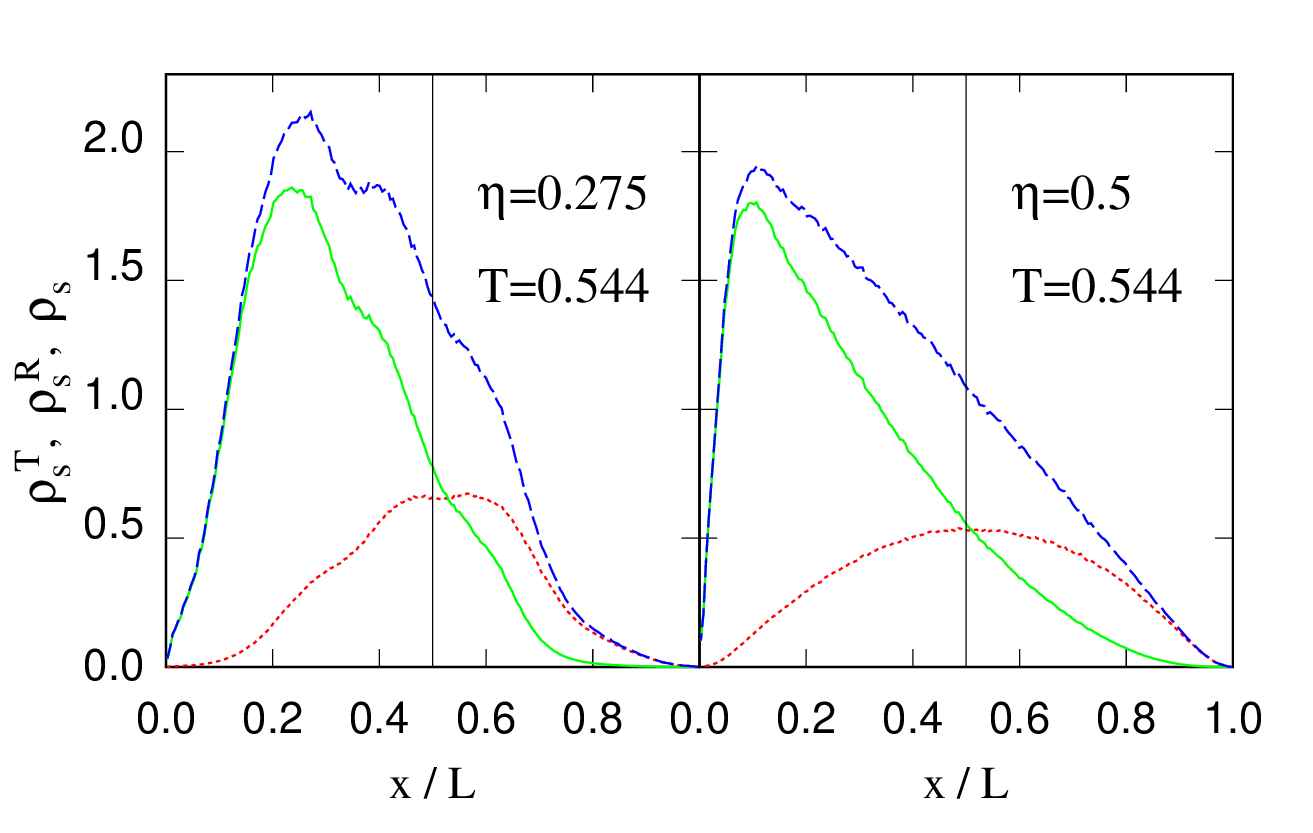}
\caption{Probability density of scattering events across the slab whose particles interact via a hc plus attractive tail potential. Red dash 
line: $\rho_s^T(x)$; green full line: $\rho_s^R(x)$; blue dash line: $\rho_s(x)=\rho_s^T(x)+ \rho_s^R(x)$. The results for panel (a) have 
been obtained for a sample close to the critical state of the corresponding homogeneous system.
}
\label{rhosw}
\end{center}
\end{minipage}
\end{figure}

Simulation, in particular, have been carried out for the system having $\delta=0.5 \sigma$ (see Eq.~\ref{fofr}) with $\epsilon=1$ being the
unit of energy. According to Ref.~\cite{well}, the critical point of this model is located at $\eta^{\ast}=0.2751$ and 
$T^{ast}/\epsilon=0.5546$. Since exploring the entire phase diagram is time consuming and possibly not very informative, a series of 
simulations at the same $\eta$ values of the hc system have been carried out following the critical isotherm $T^{ast}/\epsilon=0.5546$. In 
this way, the system is always fluid and homogeneous, but in a narrow range around the critical density, it explores in a continuous way the 
change from a gas/vapour to a liquid state.

It turns out that heavily averaged properties of the slab such as the transmission and reflection coefficients are not greatly changed by the 
addition of the attractive tail, but characteristic and systematic variations are apparent around the critical density at $\eta^{\ast}$ 
(see the comparison in Fig.~S4 of Ref.~\cite{supple}). However, but properties more dependent on particle-particle correlations are 
significantly affected. Already the average length of the transmitted light rays shows important deviations from the hard disk result (see 
Fig.~\ref{al}). As expected, differences are relatively minor at low $\eta$, where correlations are weak in both cases. Moreover, the two 
sets of results converge again at high $\eta \geq 0.4$, since the structure is dominated by packing effects, common to the two models. Large 
differences, however, are apparent at intermediate $\eta$'s where density fluctuations related to the critical state dominate the picture in 
the hc plus attractive tail case. Paths in the slab with attractive colloids, in particular, are markedly shorter than those in the slab with 
purely hc particles. The effect on the reflected paths is more uncertain, since the estimate of the $\langle l \rangle^R$ path length and
of its variance fluctuate much more that the corresponding quantities measured on transmitted paths. The onset of critical opalescence is
certainly not apparent in the simulation, but in experiments it is also known to occur very close to criticality, while the critical point
might be much less well defined in the small, inhomogeneous systems simulated in the present study. Moreover, critical opalescence might 
crucially depend on fluctuations whose wavelength is much longer that the side of the simulated cell.

The difference in $\langle l \rangle$ for the two model potentials, of course, reflects a correspondingly sizeable difference in the 
probability distribution of lengths $P_T(l)$, whose comparison is shown if Fig.~S5 of Ref.~\cite{supple}. In both cases $P_T(l)$ shows 
peculiar shapes at low $\eta$, and 
is faithfully reproduced by the inverse Gaussian distribution at medium and high $\eta$. Besides the difference in $\langle l \rangle$, around
the critical density, the $P_T(l)$ for the hc plus tail samples are systematically characterised by significantly lower variance and by 
slightly reduced skewness and kurtosis than the curve for purely hc systems.

\begin{figure}[!htb]
\begin{minipage}[c]{\textwidth}
\vskip 0.7truecm
\begin{center}
\includegraphics[scale=0.60]{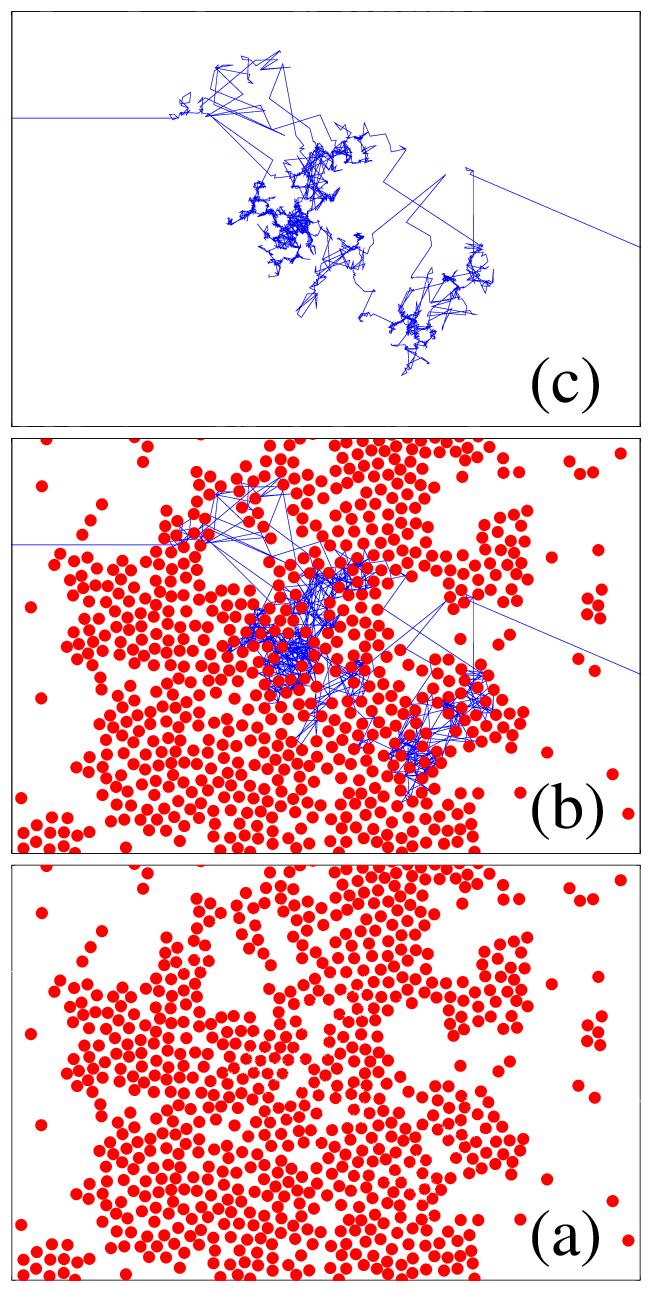}
\caption{(a) Snapshot of the 2D sample of hard disks (red dots) plus the attractive square well tail (see text) at conditions ($\eta=0.3$, 
$T/\epsilon=0.5546$) close to criticality. (b) Path of a ray of light (blue line) in the sample of panel (a); (c) the scattering disks have 
been removed to highlight the different properties of light propagation in dense and dilute domains.
}
\label{pathKrit}
\end{center}
\end{minipage}
\end{figure}

Even more apparent is the difference between the two systems in the spatial probability distribution of scattering events. While the results
for the hard disk samples display marked regularity and are easy to interpret, those for the sample of particles with the attractive tail 
display apparent irregularities at densities relatively close to the critical point (see Fig.~\ref{rhosw} (a)). Detailed analysis of 
trajectories show that the irregularities in $\rho^T_s$, $\rho^R_s$, $\rho_s$ arise from the heterogeneous spatial distribution of scattering 
centres, and of the length of straight path segments in between scattering events, as shown in Fig.~\ref{pathKrit}. It is likely that these 
irregularities would disappear upon accumulating statistics over exceedingly long production runs, but the present simulations are already 
very long, and the irregularities are a clear sign that, as expected, the time to achieve equilibrium grows substantially in approaching the 
critical point, although  any divergence is prevented by the finite (and relatively small) size of the sample. At significantly lower or 
higher packing, the probability distribution of scattering events resembles that of the slab with purely hc particles (see Fig.~\ref{rhosw} 
(b)). 

It might be useful to remark that, as suggested by Fig.~\ref{pathKrit}, both the fluctuations at all length scales as well as the {\it 
lacunarity} (using the terminology of Ref.~\cite{venice}) of samples close to criticality are qualitatively reminiscent of those of 
fluids in fractal dimension (see experimental studies in Ref.~\cite{fat, bert} as well as Ref.~\cite{loewen} for an example of 
computational fractal structure based on particles).
Therefore, the multiple-scale character of light trajectories in critical systems might also qualitatively resemble those propagating in
fractal systems. A generalisation of the present study to fractal dimensions, however, would be confronted to the extension of
the geometric (elastic) scattering rules to particles of the same or different fractal dimension. Multiple-scale heterogeneities in the
structure of the scattering centres, reflected in equally heterogeneous light trajectories, can be found in a broader class of systems,
see Ref.~\cite{marshak} and Ref.~\cite{car}.

\subsection{The 3D simulations}
\label{3D}

To gain insight into the interplay of light scattering with dimensionality and inter-particle correlations, the computations carried out in 
2D have been repeated for 3D systems. Both in 2D and in 3D the samples are labelled by a $\eta$ parameter whose definition is different but 
analogous in the two dimensionalities. However, using only $\eta$ to establish a correspondence between 2D and 3D is somewhat incomplete, 
since the same
sequence of $\eta$ values spans very different sample side $L$ or number of particle $N$ ranges in the two cases, affecting the comparison
of transmission, reflectivity, average path length, interfacial effects, etc. For instance, using in 3D the same number of particles $N=1034$ of the 2D systems, at equal $\eta$ gives 3D samples whose side $L$ is much shorter than in 2D. Being impossible to balance all aspects, as a 
compromise, in 3D we adopt a somewhat larger number of particles $N=5039$. Then, the sides $L_x=L_y=L_z(\equiv L)$ of the 3D slabs have been 
determined in order to obtain the same sequence $\eta \in [0.01; 0.05\times j; \ j=1, ..., 10]$ considered in 2D. In this way, the range of 
2D and 3D sides $L$ partially overlap, although the two sequences differ from each other.

\begin{figure}[!htb]
\begin{minipage}[c]{\textwidth}
\vskip 0.7truecm
\begin{center}
\includegraphics[scale=0.60]{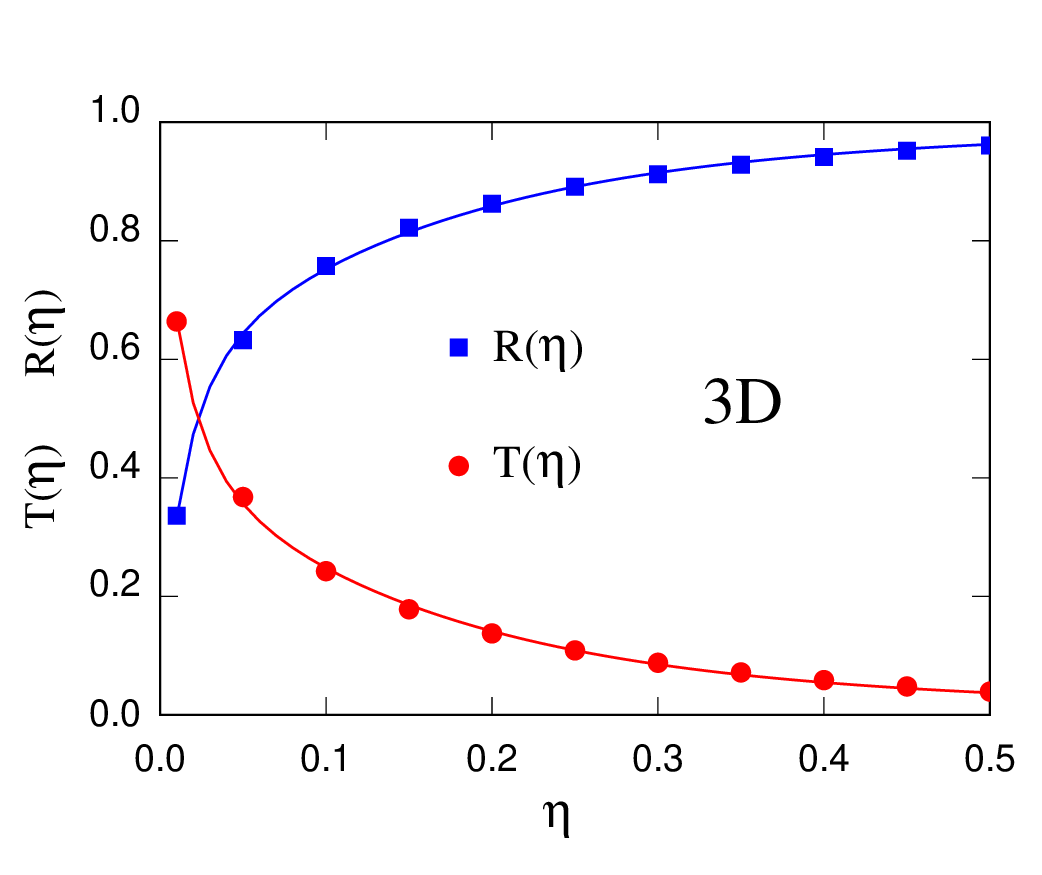}
\caption{Transmission ($T$, red dots) and reflection ($R$ blue squares) coefficients for the 3D slabs of purely repulsive hc particles as a 
function of packing fraction $\eta$. The full lines are the fit described by Eq.~\ref{pade}.
The number of particles is the same in all samples, therefore, the linear size of the slab decreasing with increasing $\eta$ like $L \propto
\eta^{-1/3}$
}
\label{rt3D}
\end{center}
\end{minipage}
\end{figure}

The transmission and reflection coefficient computed for the 3D samples are shown in Fig.~\ref{rt3D}. For the reasons discussed above, these
data cannot be quantitatively compared to those of 2D systems. However, it is apparent that the simple Pad{\'e} form of Eq.~\ref{pade} (with 
re-optimised coefficients) provides an equally good fit of the 2D and 3D results.

\begin{figure}[!htb]
\begin{minipage}[c]{\textwidth}
\vskip 0.7truecm
\begin{center}
\includegraphics[scale=0.60]{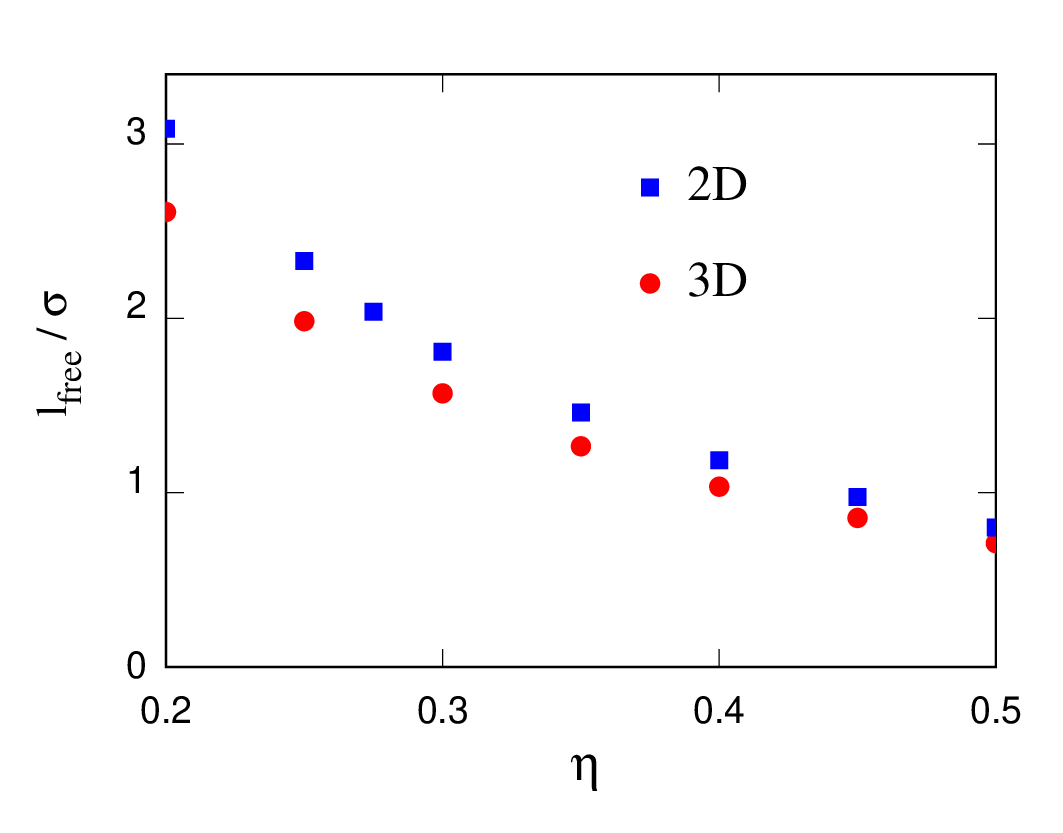}
\caption{Comparison of the mean free path in 2D and in 3D
}
\label{free3D}
\end{center}
\end{minipage}
\end{figure}

The parameters characterizing the probability distribution of transmitted path lengths $P_T(l)$ in 3D are summarised in Tab.~\ref{stat3D}. At 
variance from the 2D case, the minimum in the average path length $\langle l \rangle$ as a function of $\eta$ is not seen in the results of 
the 3D simulations, but the argument underlying its existence is valid also in 3D, and the trends in the computed data simply imply that the 
minimum occurs around or below the minimum packing fraction $\eta=0.01$ simulated in the present study. Apart from this aspect, all the other 
trends are qualitatively the same in 3D as in 2D. On the 3D data, the effect of the $\delta$-like peak in $P_T(l)$ at $l=L_x$ due to path 
crossing without being scattered is significant up to $\eta=0.15$, apparently because of the relatively limited thickness of the slab. 
Neglecting this peak, at medium-high $\eta$ ($\geq 0.15$) the inverse Gaussian distribution provides an excellent fit of the smooth part 
$\bar{P}_T(l)$ of $P_T(l)$, while the fit is poor at lower $\eta$ (see Fig.~S6  and Fig.~S7 in Ref.~\cite{supple}), presumably for the 
same reasons discussed 
for the 2D case, i.e., at low $\eta$ the propagation of light is not diffusive. The length distribution for reflected rays in 3D displays the 
same qualitative features of the 2D case. The main quantitative parameters are given in Tab.~\ref{tabR3D}.

The mean free path of rays in the colloidal suspension is the most local property of the light paths diffusing in the 
slab, therefore it could be compared directly for the two dimensionalities. The comparison between the mean free path in 2D and 3D simulations
is displayed in Fig.~\ref{free3D} for $0.2 \leq \eta \leq 0.5$. Lower values of $\eta$ have been excluded because the contribution from 
unscattered rays is sizeable and it can obscure the comparison of the diffusive contribution. One could expect that the mean free path is 
shorter in 2D, since localization by disorder is enhanced by reducing dimensionality \cite{liccia}. Somewhat surprisingly, the mean free path 
is slightly but systematically higher in 2D samples than in the 3D ones. 

\begin{table}
 \begin{tabular}{||c| c| c| c| c| c| c| c| c| c| c| c||}
  \hline
 \multicolumn{12}{|c|}{Transmitted Path Length Distribution: Statistical Moments} \\
 \hline
 \textbf{$\eta$} & \textbf{0.01} & \textbf{0.05} & \textbf{0.1} & \textbf{0.15} & \textbf{0.2} & \textbf{0.25} & \textbf{0.3} & \textbf{0.35} & \textbf{0.4} & \textbf{0.45} & \textbf{0.5}\\ [0.5ex] 
 \hline\hline
 \textbf{Mean }  & 91    & 98       & 111   & 124   & 138   &  153  & 168   & 186    & 206    &  228   &  253   \\ 
                 & (-)   & (-)      & (-)   & (154) & (159) & (169) & (182) & (198)  & (217)  & (237)  & (263)  \\ 
 \hline
 \textbf{SDev}   & 57    & 69       & 74    &  81   & 89    &  97   & 107   & 118    & 131    &  144   & 161    \\
                 & (-)   & (-)      & (-)   & (138) & (128) & (129) & (132) & (140)  & (151)  & (163)  & (178)  \\ 
 \hline
\textbf{Skew}    & 3.5   & 2.1      & 1.9   & 1.8   & 1.8   & 1.8   & 1.8   & 1.8    & 1.8    &  1.8   &  1.8   \\
                 & (-)   & (-)      & (-)   & (3.2) & (2.8) & (2.6) & (2.5) & (2.4)  & (2.3)  & (2.2)  & (2.2)  \\ 
\hline
 \textbf{ExcKur} & 16.7  & 6.2     &   5.4  &  5.2  &  5.2  &  5.1  &  5.2  &  5.2   &  5.2   &  5.3   &  5.4   \\
                 & (-)   & (-)     &  (-)   & (17.6)& (13.4)& (11.4)& (10.1)& (9.2)  & (8.7)  & (8.3)  & (7.9)  \\ 
 \hline
 \hline
\end{tabular}
\caption{Mean value (Mean), standard deviation (SDev), skewness (Skew) and excess kurtosis (ExKur) of the probability distribution $P(l)$ for the 
length $l$ of transmitted paths in 3D samples of hard spheres as a function of the packing fraction $\eta$. The corresponding values
obtained by the fit of the continuous part of $P(l)$ with an inverse Gaussian distribution  are listed in parentheses.
}
\label{stat3D}
\end{table}

The dependence of the average number of scattering events experienced by transmitted rays and estimated as $n_s=\langle l\rangle / l_{free}$ 
is similar to that already discussed for 2D, i.e., the growth is slow at low $\eta$, and becomes faster with increasing $\eta$. This aspect, 
however, might be influenced by the dependence of the scattering properties on the slab thickness $L_x$.

\begin{figure}[!htb]
\begin{minipage}[c]{\textwidth}
\vskip 0.7truecm
\begin{center}
\includegraphics[scale=0.60]{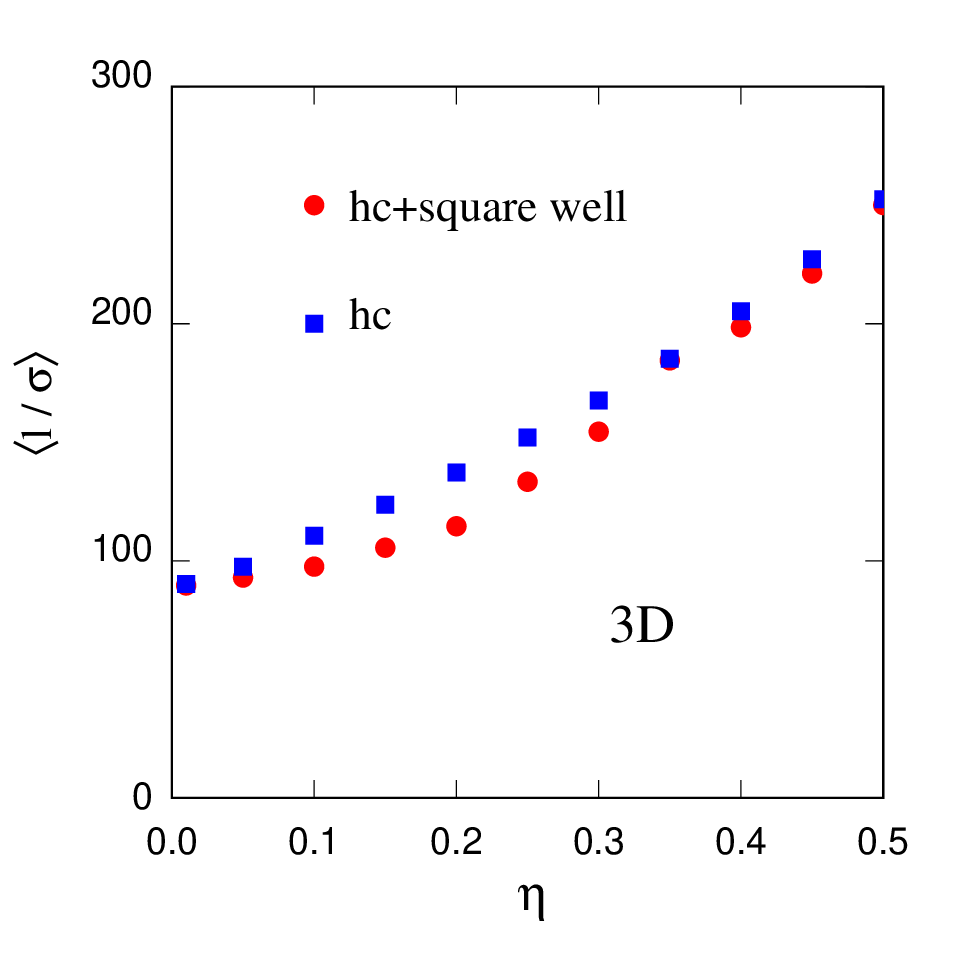}
\caption{Average length $\langle l \rangle$ of transmitted paths as a function of $\eta$. Blue filled squares: hs potential; red dots: hs 
plus attractive square well tail. 
}
\label{path3D}
\end{center}
\end{minipage}
\end{figure}

The density distribution of scattering events along the $x$-coordinate in 3D systems is analogous to the one in 2D in virtually every
aspect, including the near-symmetry of $\rho_s^T(x)$ around $L_x/2$, the linearity of the total $\rho_s(x)$ and its anomalies at low $\eta$.
These features are illustrated in Fig.~S8 of Ref.~\cite{supple}.

The effect of the interfaces, and of the density pile up at $x=0$ and $x=L_x/2$, in particular, has been assessed again by considering a
homogeneous system in which pbc are applied to all directions. It has been verified that at high $\eta$ the effect of interfaces is comparable
in 2D and in 3D.

\begin{table}
 \begin{tabular}{||c| c| c| c| c| c| c| c| c| c| c| c||}
  \hline
 \multicolumn{12}{|c|}{Reflected Path Length Distribution: Statistical Moments} \\
 \hline
 \textbf{$\eta$} & \textbf{0.01} & \textbf{0.05} & \textbf{0.1} & \textbf{0.15} & \textbf{0.2} & \textbf{0.25} & \textbf{0.3} & \textbf{0.35} & \textbf{0.4} & \textbf{0.45} & \textbf{0.5}\\ [0.5ex] 
 \hline\hline
 \textbf{Mean } & 103 &  64      &  50   &  45   &  40  &  35   &  33     &   31   & 29     &  28    &  27     \\ 
 \hline
 \textbf{SDev} &  83 &  70       &   62  &  60   &  60   &  60   &  63     &  64    &  66    &  68    &  71     \\
 \hline
\end{tabular}
\caption{Mean value (Mean) and standard deviation (SDev) of the probability distribution $P_R(l)$ for the length $l$ of reflected paths in 3D 
samples of hard spheres as a function of the packing fraction $\eta$. Statistical error bars are implicitly indicate by the number of digits of the data.
}
\label{tabR3D}
\end{table}

Dimensionality is a crucial parameter in the physics of critical states. For this reason, the comparison of properties for 3D hard spheres 
and hard spheres plus an attractive tail has been carried out in full analogy with the 2D case, following again the critical isotherm. The 
definition of the potential model is the trivial extension of the 2D case Eq.~\ref{fofr}, and the same $\delta=\sigma/2$ has been used in the 
two cases. The critical properties of the model are determined and discussed in Ref.~\cite{krit3D}. According to this reference, the 
critical point of the model occurs at $\eta^{\ast}=0.157 \pm 0.01$ and $T^{\ast}/\epsilon=1.219 \pm 0.008$. Our samples differ from those used in Ref.~\cite{krit3D} because they comprise a larger number of particles, and because they are limited by two parallel interfaces. 
Due to these differences, the sample at exactly the critical parameters of Ref.~\cite{krit3D} appears to be divided in two slightly 
different phases. To make sure that the sample is homogeneous, the temperature of the simulation has been changed slightly to 
$T/\epsilon=1.25$. At this temperature, the MC results still display large density fluctuations at $\eta$ close to $\eta^{\ast}$ that 
persist over long times apparently because of critical slowing down.

The results for the average length $\langle l \rangle$ of paths transmitted through the suspension of square well particle is shown in 
Fig.~\ref{path3D}, compared with the results of the 3D hard sphere model. The trend is the same of the 2D case, i.e., close to criticality
$\langle l \rangle$ is lower for the hard sphere plus attractive tail model than for hard sphere samples, but the quantitative difference 
between the two sets of results is somewhat smaller in 3D, consistently with the generally accepted statement that the effects of criticality 
are stronger in 2D than in 3D. Also in this 3D case, the average length of reflected beams is longer for particles with the attractive tail
that for pure hard spheres.

\subsection{Amplification and output power distribution}
\label{ampll}

\begin{figure}[!htb]
\begin{minipage}[c]{\textwidth}
\vskip 0.7truecm
\begin{center}
\includegraphics[scale=0.60]{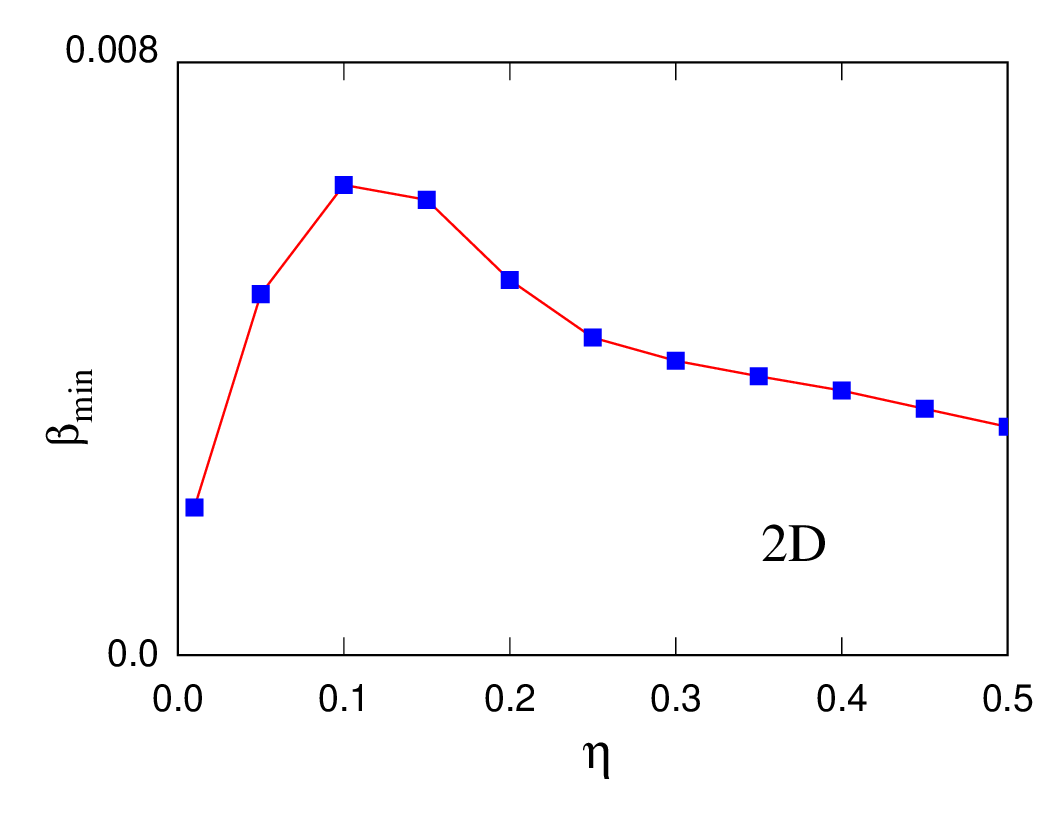}
\caption{Blue filled squares: Minimum value of the gain per unit length $\beta$ required to achieve amplification $k_{ampl}>1$. The line is 
a guide for the eye.
}
\label{amply}
\end{center}
\end{minipage}
\end{figure}

In random lasers based on suspensions of colloidal particles, scattering and diffusion of light underlie the amplification of the input beam
through stimulated emission in the solution. Since the model discussed in the previous sections contains no detail on the optical properties 
of the medium, the discussion of lasing properties is limited to statistical considerations on the length of transmitted paths and on the
intensity of the emitted rays. 

Each transmitted ray of initial intensity $i_0\equiv I_{inp}$ (ideally, entering the slab as a single photon) will leave the slab after 
covering a distance $l$ in its diffusive drift, emerging with an output intensity $I_{out}=i_0 \ {exp}{(\beta l)}$, where $\beta$ is the 
gain per unit length due to stimulated emission (see Sec.~\ref{method}). The overall amplification factor $k_{ampl}$, therefore will be:
\begin{equation}
k_{ampl}=\langle \frac{I_{out}}{I_{inp}}\rangle=\frac{ \sum_{i\in T}\ {exp}{(\beta l_i)}}{N_T+N_R}
\end{equation}
where $\sum_{i\in T}$ indicates the sum over the set of transmitted rays, while $N_T$ and $N_R$ are the number of transmitted and reflected 
rays, respectively. Provided $\beta \langle l \rangle$ is small, $k_{ampl}$ can be approximated as:
\begin{equation}
k_{ampl}=T \exp{(\beta \langle l \rangle)}
\end{equation}
(as anticipated in Eq.~\ref{aaampl})
where $T$ is the transmission coefficient and $\langle l \rangle$ is the average of the individual transmitted path lengths $\{ l_i\}$. More 
precisely, because of the inequality of arithmetic and geometric means \cite{amgm}, the relation between the two expressions for $k_{ampl}$ is 
in fact an inequality:
\begin{equation}
k_{ampl}=\frac{I_{out}}{I_{inp}}=\frac{ \sum_{i\in T}\ \exp{(\beta l_i)}}{N_T+N_R}> T\ \exp{(\beta \langle l \rangle)}
\label{suffy}
\end{equation}
that becomes an equality in the limit of vanishing $\beta \langle l \rangle$:
For this reason, $T\exp{(\beta \langle l \rangle)}>1$ is a sufficient condition for amplification. Using the simulation data for $T$,
it is possible to determine for each $\eta$ the minimum $\beta$ that provides amplification. The results are shown in Fig.~\ref{amply}
for 2D and in Fig.~S9 of Ref.~\cite{supple} for 3D, and represent the computational analogue of the threshold dependence on dye 
molecules and scatterer concentration highlighted in experiments \cite{depend}.
It might be useful to point out that the results strictly depend on the size of the slab. It is precisely because of 
the ambiguous comparison of size in 2D and in 3D that the results for the two different dimensionalities have not been reported on the same
plot. The results show that the optimal range is at low and at high density, while the medium density range is somewhat disfavoured, since
it requires larger values of $\beta$ to achieve amplification.

\begin{figure}[!htb]
\begin{minipage}[c]{\textwidth}
\vskip 0.7truecm
\begin{center}
\includegraphics[scale=0.60]{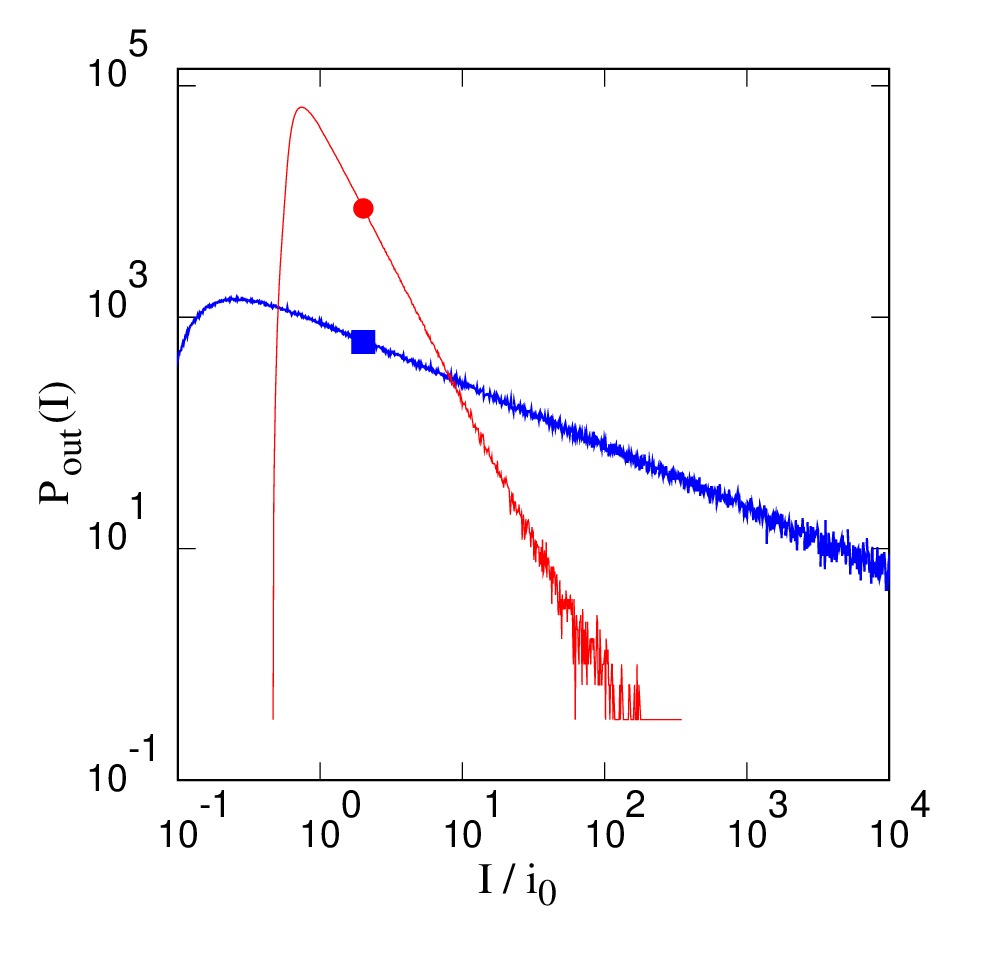}
\caption{Probability distribution of the output intensity I in 2D. Red curve (and dot): $\eta=0.15$; blue curve (and filled square): 
$\eta=0.5$. The output has been computed assuming the arbitrary value $\beta=0.002$ as the amplification per unit length.
}
\label{Lev}
\end{center}
\end{minipage}
\end{figure}

Besides the overall amplification, which is of interest for lasing, an important issue concerns the statistics of the power output from the 
slab made of colloidal particles in an optically active medium. The probability distribution $P_{out}(I)$ of the ensemble of output 
intensities has peculiar properties in the limit of large $I$, that have been extensively investigate in the past for their conceptual and 
applied interest \cite{fat}.

As already stated, in the simple picture underlying the present study, a fraction $T$ of the incoming rays will exit the slab with an 
intensity $I_i$ amplified by a factor that depends on the length $l_i$ of their trajectory within the slab. Then, the simulation data for the 
length of each transmitted ray allows the direct computation of the intensity distribution of the emitted rays. The result for the high 
intensity range is reported in Fig.~\ref{Lev} for the 2D case with $\eta=0.1$ and $\eta=0.5$. The 
double logarithmic scale makes it apparent that the emitted intensity follows an inverse power law behaviour, whose exponent depends on the 
parameter $\beta$ selected for estimating the amplification. The same behaviour is observed in all simulated cases, both in 2D and in 3D 
samples. In experiments, inverse power law decay of $P_{out}$ usually is associated with a L{\'e}vy distribution, but its observation is far 
less general than in simulation \cite{levy}. It is likely that the difference between simulation and experiment is due to the simplicity of the
model, that neglects every feature related to the microscopic optical processes underlying stimulated emission. In any case, the computational
results depend on the $\beta$ value, which unfortunately is arbitrary.

Additional insight into this statistical mechanics aspect can be gained from the observation that both in 2D and in 3D the Wald distribution 
gives a good fit of the smooth part $\bar{P}_T(l)$ of the distribution of transmitted path lengths, provided $\eta$ is not too small.
In the present model, a transmitted ray of initial intensity 
$i_0$ is emitted with a final intensity multiplied by $\exp{(\beta l)}$, therefore, the probability distribution $P_{out}(I)$ of emitted 
intensities $I$ is fully determined by $P_T(l)$, since rays of intensity $I$ are emitted by all paths whose length is:
\begin{equation}
l(I)=\frac{\ln{I}}{\beta}-\ln{(i_0)}
\end{equation}
which, at high $I$ is approximated as $l(I)=\ln{I}/\beta$. However, the probability of paths having length $l >> \langle l \rangle$ is 
$\bar{P}_T(l)$. 
Hence, using:
\begin{equation}
P_{out}(I)=\bar{P}_T(l(I)) \frac{dl(I)}{dI}
\end{equation}
one obtains:
\begin{equation}
P_{out}(i)=\frac{1}{I\beta}\sqrt{\frac{\lambda \beta^3}{2\pi (\ln{I})^3}} \exp{\left[ -\frac{\lambda}{2\beta} 
\frac{(\ln{I}-\mu \beta)^2}{\mu^2 \ln{I}}\right]}=
\end{equation}
\begin{displaymath}
=\frac{1}{I\beta}\left(\frac{1}{I^{1/2}}\right)^{\frac{\lambda}{\beta\mu^2}}\sqrt{\frac{\lambda \beta^3}{2\pi (\ln{I})^3}}
\times C\times \exp{\left[ -\frac{\lambda \beta}{\ln{I}}\right]}
\end{displaymath}
where $C=\exp{(\lambda/\mu)}$ is a constant.

In the limit of high power $I$, the last exponential tends to 1. Neglecting the dependence of $P_{out}$ on the logarithmic factor 
$(\ln{I})^{-3/2}$ (whose derivative tends to zero in the limit of large $I$), the leading dependence on $I$ is:
\begin{equation}
P_{out}(I)\propto \frac{1}{I}\left(\frac{1}{I^{1/2}}\right)^{\frac{\lambda}{\beta\mu^2}}
\end{equation}
which has apparent similarities with the asymptotic behaviour of the L{\'e}vy distribution. A precise equivalence requires {\it ad hoc} 
assumptions that are not inherent aspects of the model. Nevertheless, we observe that inserting into these expressions the values of $\mu$ and
$\sigma$ obtained from the fit, an the $\beta$ estimated for the lasing threshold (see Fig.~\ref{amply}), $\lambda/\beta\mu^2$ is close to 
one, and the exponent of $I$ is close to $-3/2$. Even without any assumption, the  inverse power behaviour of $P_{out}(I)$ at large $I$ 
represents a fat tail \cite{stats, fat} that seems to be a characteristic feature of the power spectrum of random lasers.
Previous studies have also attributed L{\'e}vy statistics to near threshold conditions \cite{muju}.

\section{Discussion and summary}
The propagation of light in a random medium is a subject of great conceptual ad practical interest, underlying, for instance, the development
of random lasers and applications that range from speckle-free illumination and imaging \cite{speck}, nanotechnology \cite{nano} and medical 
diagnostics \cite{human}.
Previous studies have investigated the statistical mechanics aspects of light propagation and amplification in random media using a variety
of models of different sophistication. The most representative models, however, consist of random walks among scattering centres confined
in between two parallel surfaces located at $x=0$ and $x=L_x$, leaving unconstrained the diffusion along the other directions \cite{rogers}.
In these model, each step is determined by parameters drawn from pre-assigned probability distributions, deciding the length of the step
(whose average is the mean-free path) as well as the change of propagation direction upon each scattering event. In virtually all cases these
models, although sophisticated and useful, are generic (and thus also general), in the sense that they are not derived from a specific model
of the distribution of scattering centres, and sometimes do not quantitatively model a specific scattering mechanism. The novelty of the 
present approach is that it is tailored on a specific model of colloidal particles in a fluid. Correlations among particles are explicitly 
accounted for, and the effect of the interfaces, of different inter-particle potentials or different thermodynamic conditions arise from the 
underlying model of particles and interfaces. Admittedly, the scattering mechanism implemented in the present model is simple, consisting
of ideal reflection at the surface of the colloidal particles, and the wave character of light is neglected. These aspects, however, 
could be refined at least to some extent, especially in the direction of introducing different scattering mechanisms.

In this broad framework, two basic many-particle models have been investigated, consisting of purely repulsive hard-core particles and of 
similarly repulsive particles whose interaction potential, however, includes a square well attractive tail. Simulations have been carried out 
in 2D and in 3D, although, for computational convenience and ease of visualisation, more effort and emphasis have been devoted to the 2D case.The 2D case is not only a theoretical playground, since virtually 2D random lasers \cite{2D} have also been made. 

Analysis of trajectories provides exact data for the propagation of light in the models that have been simulated. Transmission and reflection
coefficients have been computed in 2D and 3D, and approximated with a simple Pad{\'e} form. The average length $\langle l \rangle$ of 
transmitted paths has been computed, and found to display a massive increase at high $\eta$ with respect to the minimum length corresponding
to the width $L_x$ of the slab. At $\eta=0.5$, for instance, the average path exceeds the minimum path $L_x$ by a factor $38$ and $15$ in 2D 
and 3D, respectively. More importantly, the probability distribution $P_T(l)$ of the length $l$ of transmitted paths is broad, having high 
variance and excess kurtosis, corresponding to a sizeable number of positive ($l > \langle l \rangle$) outliers. Given the exponential 
character of the amplification by the optically active medium, these outliers may have a disproportionate effect on the output power. At 
medium-high $\eta$, the Wald (also known as inverse Gaussian) distribution provides an accurate representation of $P_T(l)$ even in the 
long-$l$ tail. Translating the analytical form of the Wald distribution into the power domain shows that the distribution of power intensity 
for the outcoming light follows an inverse poser decay on the high-intensity range. Therefore, even the simple model that has been simulated 
has a characteristic output distribution characterised by the fat tail that has been extensively discussed for random lasers \cite{fat2}.

The distribution of scattering events (see Fig.~\ref{rhos}) shows that, apart interfacial effects that are sizeable but localised, the 
intensity of light decreases linearly with increasing $x$, as a result of the diffusive propagation. Deviations are expected and observed at 
low $\eta$ when the propagation of light is partly ballistic. This observation implies a Ohmic dependence of the transmission coefficient on 
the width of the slab, that has been verified by a single computation in 2D at $\eta=0.300$ by doubling $N$ as well as the width of the 
simulated slab.

Additional simulations have been performed on slightly modified models to investigate the effect of interfaces and of weak polydispersivity
on the system properties. More importantly, comparison between the results for the purely repulsive potential and for the case with a square
well attractive tail shows that the important fluctuations around the critical state of the later model decrease the length of the transmitted
paths, and decrease the width of the $P_T(l)$ distribution, while the kurtosis is enhanced. This shows that the performance of random lasers
based on colloidal particles can be tuned by changing the interaction among particle and the thermodynamic state of the system.

An example of complex property that can be investigated directly by a detailed and specific model is the loss of memory in the direction
of rays in the diffusive regime of light propagation. Memory is rapidly lost, but on the short time scale it is sizeable and peculiar, showing
a predominant effect of backscattering. This property depends on the interplay of the scattering matrix of each reflection event with the
correlation in the particles' position, and it is difficult to model a priory. In the present approach, it emerges automatically from the
explicit description of correlations, including the effect of interfaces or of medium scale inhomogeneities. 

In the future, the scattering matrix (here introduced only implicitly through the rules of geometric reflection) of different scattering 
mechanisms could be implemented, and the spatial correlation in the distribution of scattering centres could be modeled for different types 
of disordered media, providing quantitative data for specific materials and geometries. On the other hand, a simple modification of the 
experimental setup of Ref.~\cite{viola}, probing the systems with a laser beam narrower than the mm size of the particles, could
closely reproduce the conditions of the computational model. Then, comparison of the new results with those obtained using a broad, 
plane-wave like illumination, could disentangle the geometric aspects of the light propagation from the complexities of the scattering
mechanism.

\end{document}